\documentstyle[preprint,aps,epsf]{revtex}  %% Draft Style             
\tightenlines
\begin{document}
\draft

% Include following two lines for Journal Style
%\twocolumn[\hsize\textwidth\columnwidth\hsize  %**Journal
%\csname @twocolumnfalse\endcsname              %**Journal

\title {\null\vspace*{-.0cm}\hfill {\small nucl-th/0106067} \\ \vskip
0.8cm Heavy Quarkonium Dissociation Cross Sections\\ in Relativistic Heavy-Ion
Collisions}

\author{Cheuk-Yin Wong$^1$, E. S. Swanson$^{2,3}$ and T. Barnes$^{1,4}$}

\address{$^1$Physics Division, Oak Ridge National Laboratory, Oak Ridge,
TN 37831 USA}

\address{$^2$Department of Physics and Astronomy,
University of
Pittsburgh, Pittsburgh, PA 15260 USA }

\address{$^3$Jefferson Lab, Newport News, VA 23606 USA}

\address{$^4$Department of Physics, University of Tennessee, Knoxville, TN
37996 USA}

\maketitle

\begin{abstract}
{Many of the hadron-hadron cross sections required for the study of
the dynamics of matter produced in relativistic heavy-ion collisions
can be calculated using the quark-interchange model.  Here we evaluate
the low-energy dissociation cross sections of $J/\psi$, $\psi'$,
$\chi$, $\Upsilon$, and $\Upsilon'$ in collision with $\pi$, $\rho$,
and $K$, which are important for the interpretation of
heavy-quarkonium suppression as a signature for the quark gluon
plasma.  These comover dissociation processes also contribute to
heavy-quarkonium suppression, and must be understood and incorporated
in simulations of heavy-ion collisions before QGP formation can be
established through this signature.}

\end{abstract}

\newpage

\section{Introduction}

The first collisions of heavy-ion beams at the Relativistic Heavy-Ion
Collider (RHIC) at Brookhaven National Laboratory heralded a new era
in the study of matter in the extreme conditions of very high
temperatures and energy densities.  The ultimate goal of this research
is the production of a quark-gluon plasma, which is an unusual phase
of strongly interacting matter that purportedly existed shortly after
the Big Bang \cite{QM99,Won94}.

The search for the quark-gluon plasma relies on the unusual properties
of the plasma for its detection.  However, many conventional hadrons
are also produced during a heavy ion collisions.  Whatever signal is
chosen for the identification of the quark-gluon plasma, contributions
to that signal from conventional hadronic processes must be identified
as backgrounds and removed from the data.

Although there are recent, tantalizing hints of possible quark-gluon
plasma production in heavy-ion collisions at CERN \cite{Hei00,Won01},
conclusive evidence is still lacking due to uncertain backgrounds from
conventional hadronic sources.  Investigations of the various
`hadronic background processes' are urgently needed if we are to
develop a satisfactory understanding of the various signals proposed
as signatures of a quark-gluon plasma.  In this paper we consider one
type of hadronic background processes that can contribute to
heavy-quarkonium suppression, which is frequently cited as a QGP
signature.

Matsui and Satz \cite{Mat86} originally suggested the use of
suppressed $J/\psi$ production as a signature for the formation of a
quark-gluon plasma in high-energy heavy-ion collisions.  The recent
experimental observation of anomalous $J/\psi$ suppression in Pb+Pb
collisions by the NA50 Collaboration \cite{Gon96,Rom98} has been
considered by many authors
\cite{Won96,Won98,Kha96,Bla96,Cap96,Cas96,Vog98,Nar98,Zha00,Bla00}.
However there is considerable uncertainty in these studies, due to the
lack of reliable experimental information on $J/\psi$ and $\chi_J$
dissociation cross sections in low-energy collisions with light
hadrons.  Because heavy quarkonia decay strongly, many of the
dissociation cross sections cannot be measured directly in hadron
scattering experiments; the cross sections are instead typically
estimated using theoretical models.  Evaluation of these cross
sections is of particular interest for clarifying the physics of the
$J/\psi$ anomalies observed in Pb-Pb collisions, and may be of
considerable importance in future $J/\psi$ studies using the RHIC and
LHC colliders.

The dissociation of the $J/\psi$ by hadrons has been considered
previously in several theoretical studies, but the predicted cross
sections show great variation at low energies, largely due to
different assumptions regarding the dominant scattering mechanism
\cite{Kha94,Kha96a,Mar95,Won00a,Won00b,Bar00,Mat98,Hag00,Lin00,Oh00,Sib00}.

Kharzeev, Satz, and collaborators \cite{Kha94,Kha96a} employed the
parton model and perturbative QCD ``short-distance'' approach of
Bhanot and Peskin \cite{Pes79,Bha79}, and found remarkably small
low-energy cross sections for collisions of $J/\psi$ with light
hadrons.  For example, their $J/\psi$+$ N$ cross section at
$\sqrt{s}=5$~GeV is only about 0.25~$\mu b$ \cite{Kha94}. A
finite-mass correction increases this cross section by about a factor
of two \cite{Kha96a}.  However, in high-energy heavy-ion reactions the
collisions between the produced $\pi$ and $\rho$ with $J/\psi$ and
$\psi'$ occur at low energies (typically from a few hundred MeV to
about 1~GeV relative kinetic energies). The applicability of the
parton model and pQCD for reactions in this low-energy region is
certainly open to question.

Matinyan and M\"uller \cite{Mat98}, Haglin\cite{Hag00}, Lin and
Ko\cite{Lin00}, and Oh, Song, and Lee \cite{Oh00} recently reported
results for these dissociation cross sections in meson exchange
models.  These references all use effective meson Lagrangians, but
differ in the interaction terms included in the Lagrangian.  Matinyan
and M\"uller included only $t$-channel $D$ meson exchange, and found
that the dissociation cross sections of $J/\psi$ by $\pi$ and $\rho$
are rather small; both are $\approx 0.2$-$0.3$~mb at $\sqrt{s}=4$~GeV.
Including form factors (arbitrarily chosen to be Gaussians with a
width set to 1.5 GeV) would reduce these cross sections by an order of
magnitude. Haglin obtained a very different result, with much larger
cross sections, by treating the $D$ and $D^*$ mesons as non-Abelian
gauge bosons in a minimally coupled Yang-Mills meson Lagrangian. Form
factors were also introduced in these calculations
\cite{Hag00,Lin00,Oh00}.  The resulting mb-scale cross sections are
very sensitive to the choice of form factors.  Charmonium dissociation
by nucleons has also been considered recently using a similar
effective Lagrangian formulation \cite{Sib00}.  Of course the use of a
Yang-Mills Lagrangian for charmed mesons has no {\it a priori}
justification, so this crucial initial assumption made in these
references requires independent confirmation.  The assumption of the
$t$-channel exchange of a heavy meson such as a $D$ or $D^*$ between a
hadron and a $J/\psi$ is also difficult to justify physically, because
the range of these exchanges ($1/M \approx 0.1$ fm) is much smaller
than the physical extent of the initial hadron and the $J/\psi$.

Charmonium dissociation processes can undoubtedly be described in
terms of interquark interactions, as we attempt in this paper.  Since
these reactions are of greatest phenomenological interest at energy
scales in the resonance region, we advocate the use of the known quark
forces to obtain the underlying scattering amplitudes from explicit
nonrelativistic quark model hadron wavefunctions of the initial and
final mesons.

Martins, Blaschke, and Quack \cite{Mar95} previously reported
dissociation cross section calculations using essentially the same
approach we describe here.  The short-distance interaction used by
these authors in particular is quite similar to the form we
employ. For the confining interaction, however, they used a simplified
color-independent Gaussian potential between quark-antiquark pairs,
rather than the now well-established linear $\bbox{\lambda}(i) \cdot
\bbox{\lambda}(j)$ form.  They found a rather large $\pi$+$J/\psi$
dissociation cross section, which reached a maximum of about 7 mb at a
center-of-mass kinetic energy $E_{KE}$ of about 0.85 GeV.  Although
our approach is very similar to that of Martins {\it et al.}, our
final numerical results differ significantly, due mainly to our
different treatments of the confining interaction.

In this paper we use the approach discussed above to evaluate the
dissociation cross sections of $J/\psi$, $\psi'$, $\chi$, $\Upsilon$,
and $\Upsilon$ by $\pi$, $\rho$, and $K$, and compare our results to
other theoretical cross sections reported in the literature. The
dissociation cross sections of $\chi_J$ mesons are of special
interest, as about 1/3 of the $J/\psi$ mesons produced in a
high-energy nucleon-nucleon collision come from the decay of $\chi$
states \cite{Ant93}.  The dissociation cross sections for $\Upsilon$
are also interesting and they have recently been estimated to be quite
small in an effective-Lagrangian meson exchange model because of large
thresholds for the dissociation of $\Upsilon$ by both $\pi$ and $\rho$
\cite{Lin01}.

We employ the Barnes-Swanson quark-interchange model
\cite{Bar92,Swa92} to evaluate these dissociation cross sections in
terms of wavefunctions and interactions at the quark level.  We use
the nonrelativistic quark potential model and its interquark
Hamiltonian to describe the underlying quark-gluon forces.  The model
parameters and quark masses are determined by the meson spectrum, so
there is little additional freedom in determining scattering
amplitudes and cross sections.  We thus implicitly incorporate the
successes of the quark model in describing the meson spectrum and many
static and dynamical properties of hadrons.  We proceed by calculating
the scattering amplitude for a given process at Born order in the
interquark Hamiltonian; the good agreement of this approach with
experimental scattering data on many low-energy reactions
\cite{Bar92,Swa92,Kpi,KN} provides strong motivation for the
application of this approach to hadron reactions in relativistic
heavy-ion collisions.  A brief summary of the present work has been
reported previously \cite{Won00a}.

This paper is organized as follows.  In Section II we summarize the
Barnes-Swanson model of quark interchange as applied to the
calculations of the dissociation cross sections.  The reaction matrix
can be described in terms of the ``prior'' or ``post'' diagrams, which
are discussed in Section III.  Section IV gives some details of the
evaluation of the spin and spatial matrix elements for the general
meson-meson scattering problem.  In Section V the spin matrix elements
are derived explicitly in terms of 9-$j$ symbols.  The evaluation of
spatial overlap integrals for the case of all $S$-wave mesons is
discussed in Section VI.  In Section VII, we evaluate the
corresponding overlap integrals for one $P$-wave meson. An accurate
determination of these matrix elements requires correspondingly
accurate bound state wavefunctions; the evaluation of these
wavefunctions is discussed in Section VIII. The numerical agreement
between the post and prior scattering formalisms is demonstrated
explicitly in Section IX, which provides a very nontrivial check of
the accuracy and internal consistency of our calculations.  Section X
presents our results for the dissociation cross section of $J/\psi$
and $\psi'$ in collision with various light mesons, and Section XI
gives the corresponding cross sections for the dissociation of
$\Upsilon$ and $\Upsilon'$.  Section XII present results for the
dissociation of $P$-wave charmonium states, the $\chi_J$ mesons, in
collision with $\pi$, $\rho$, and $K$.  Finally, we present
conclusions in Section XIII.

\section{The Model}

We shall briefly summarize the model of Barnes and Swanson for
constituent-interchange processes in the reaction
\begin{equation}  
\label{eq:1} A(12)+B(34)\rightarrow C(14)+D(32)
\end{equation}  
where $A$, $B$, $C$, and $D$ are $q\bar q$ mesons, and 1, 3, and 2, 4
label the quark and antiquark constituents respectively.  In this
meson-meson scattering problem the scattering amplitude in the ``prior
formalism" is the sum of the four quark-line diagrams of Fig.\
1. These are evaluated as overlap integrals of quark model
wavefunctions using the ``Feynman rules" given in App.\ C of
Ref.\cite{Bar92}.  This method has previously been applied
successfully to the closely related no-annihilation scattering
channels $I=2$ $\pi\pi$ \cite{Bar92}, $I=3/2$ $K\pi$ \cite {Kpi},
$I=\{0,1\}$ $S$-wave $KN$ scattering \cite{KN}, and the short-range
repulsive $NN$ interaction \cite{NN}.

The interaction $H_{ij}(r_{ij})$ between the pair of constituents $i$
and $j$ is represented by the curly line in Fig.\ 1 and is taken to be
\begin{eqnarray}
\label{eq:Hij}
H_{ij}(r_{ij})&=&{\bbox{\lambda}(i) \over 2}\cdot {\bbox{\lambda}(j)
\over 2} \left \{
V_{\rm color-Coulomb}(r_{ij})
+V_{\rm linear}(r_{ij})
+V_{\rm spin-spin}(r_{ij}) + V_{\rm con}
\right \}  \nonumber  \\
&=&{\bbox{\lambda}(i) \over 2}\cdot {\bbox{\lambda}(j) \over 2} \left \{
{\alpha_s \over r_{ij}} - {3 b \over 4} r_{ij} - {8 \pi \alpha_s \over
3 m_i m_j } \bbox{s}_i \cdot \bbox{s}_j \left ( {\sigma^3 \over
\pi^{3/2} } \right ) e^{-\sigma^2 r_{ij}^2}  
+ V_{\rm con}
\right
\},
\end{eqnarray}
where $\alpha_s$ is the strong coupling constant, $b$ is the string
tension, $m_i$ and $m_j$ are the masses of the interacting
constituents, $\sigma$ is the range parameter in the hyperfine
spin-spin interaction, and $V_{\rm con}$ is a constant.  For an
antiquark, the generator $\bbox{\lambda}/2$ is replaced by
$-\bbox{\lambda}^{T}/2$.

\vspace*{4.5cm}
\epsfxsize=300pt
\includegraphics{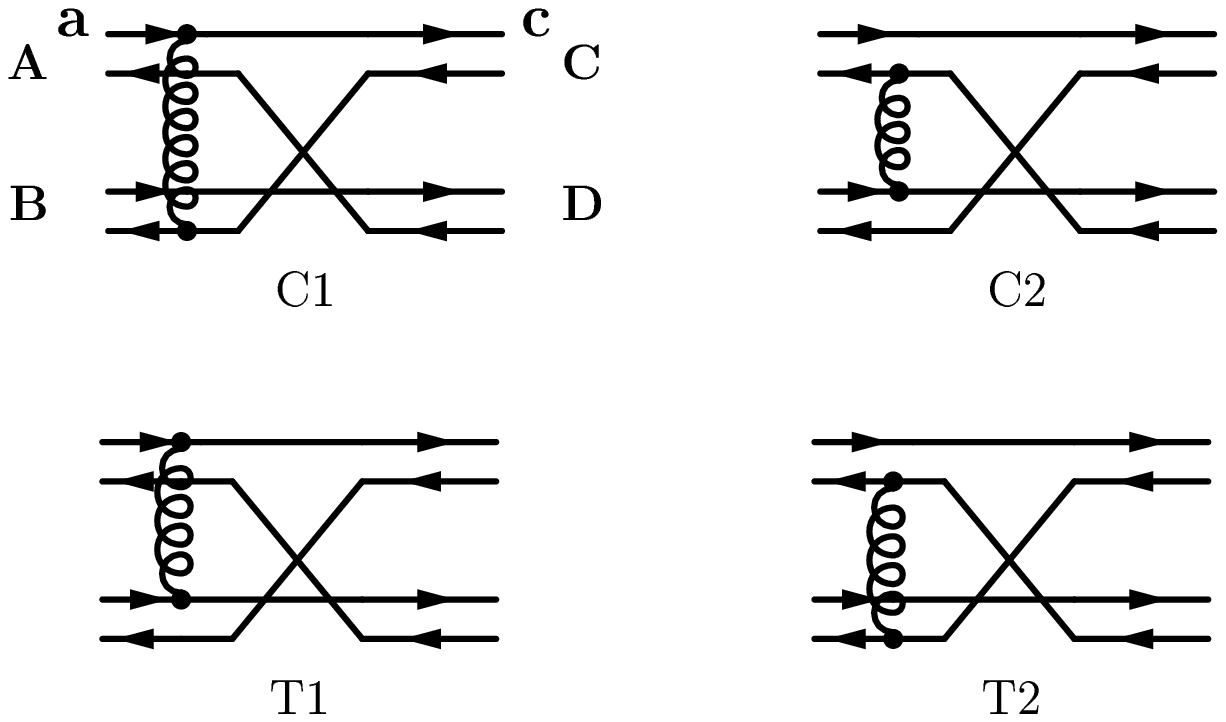}
\vspace*{0.4cm}\hspace*{2cm}
\begin{minipage}[t]{12cm}
\noindent {\bf Fig.\ 1}.  {`Prior' diagrams for Born-order meson-meson
scattering.}
\end{minipage}
\vskip 4truemm
\noindent 

It is convenient to introduce $V_{ij}(r_{ij})$ to denote the quantity
in curly brackets in Eq.\ (\ref{eq:Hij}) so that we can write
$H_{ij}(\bbox{r}_{ij})$ in the form
\begin{eqnarray}
\label{eq:vij}
H_{ij}(\bbox{r}_{ij})=
{\bbox{\lambda}(i) \over 2}\cdot {\bbox{\lambda}(j) \over 2}
~V_{ij}(r_{ij}).
\end{eqnarray}

The Born-order $T$-matrix element $T_{fi}$ is proportional to the matrix
elements $h_{fi}$ of this residual interaction (as defined in
Ref.\cite{Bar92}). For each of the scattering diagrams of Fig.\ 1,
$h_{ij}$ and $T_{ij}$ are given as the product of four factors,
\begin{eqnarray}
h_{fi} = {1\over (2\pi)^3}\, T_{fi} = S\, I_{\rm flavor}\, 
I_{\rm color}\, I_{\rm spin-space}\; .
\end{eqnarray}
The overall sign $S$ is a fermion-permutation phase known as the
``signature'' of the diagram, which is equal to $(-1)^{N_x}$, where
$N_x$ is the number of fermion line crossings. ($S=-1$ for the
diagrams in Fig.\ 1.) The flavor matrix element $I_{\rm flavor}$ is
the overlap of the initial and final flavor wavefunctions.  In all the
processes considered in this paper, $I_{\rm flavor}$ is equal to 1 for
all diagrams.  The color matrix element $I_{\rm color}$ is $-4/9$ for
diagrams C1 and C2 of Fig.\ 1, and is $+4/9$ for diagrams T1 and T2.
The spatial and spin matrix element $I_{\rm spin-space}$ is the matrix
element of $V_{ij}$, and can in general be written as a sum of
products of a spin matrix element $I_{\rm spin}$ times a spatial
matrix elements $I_{\rm space}$.  The spin matrix element $I_{\rm
spin}$ involves Clebsch-Gordon coefficients and the spins of the
colliding particles and is tabulated for all cases of $S$-wave mesons
in \cite{Bar92}.  An explicit closed-form expression for this $I_{\rm
spin}$ in terms of Wigner's 9$j$ symbols will be given in Section V.
The evaluation of the spatial matrix element $I_{\rm space}$ will be
discussed in detail in Sections VI and VII.

For the reaction $A+B \rightarrow C+D$, with an invariant momentum
transfer $t$
\begin{eqnarray}
t=(A-C)^2=m_A^2+m_C^2-2A_0C_0+2{\bbox{A}}\cdot{\bbox{C}},
\end{eqnarray}
the differential cross section is given by
\begin{eqnarray}
{d \sigma_{fi} \over dt} = { 1 \over 64 \pi s |{\bbox{p}}_A|^2 } 
\left | {\cal M}_{fi} \right |^2
\end{eqnarray}
where the matrix element ${\cal M}_{fi}$ is related to $T_{fi}$ by
\begin{eqnarray}
{\cal M}_{fi} = \sqrt{(2E_A)(2E_B) (2E_C) (2E_D)} ~~T_{fi}.
\end{eqnarray}
In Eqs. (6) and (7), ${\bbox{p}}_A$ and $E_A$ are the momentum and the
energy of meson $A$ in the center-of-mass system.  The total cross
section for the reaction $A+B \rightarrow C+D$ can be obtained from
$d\sigma_{fi}/dt$ by integrating over $t$.

\section{ Post and prior descriptions}

Before proceeding to our results, we note that a well-known
``post-prior'' ambiguity arises in the calculation of bound state
scattering amplitudes involving rearrangement collisions \cite{Sch68}.
Since the Hamiltonian which describes the scattering process $AB
\rightarrow CD$ can be separated into an unperturbed Hamiltonian and a
residual interaction in two ways, $H = H^{(0)}_A + H^{(0)}_B +
V_{AB}=H^{(0)}_C + H^{(0)}_D + V_{CD}$, there is an ambiguity in the
choice of $V_{AB}$ or $V_{CD}$ as the residual interaction.  The first
version gives the ``prior" diagrams of Fig.\ 1, in which the
interaction occurs before constituent interchange.  The second choice
is the ``post" formalism in which the interaction occurs after
constituent interchange, as in the diagrams of Fig.\ 2.

\vspace*{4.5cm}
\epsfxsize=300pt
\includegraphics{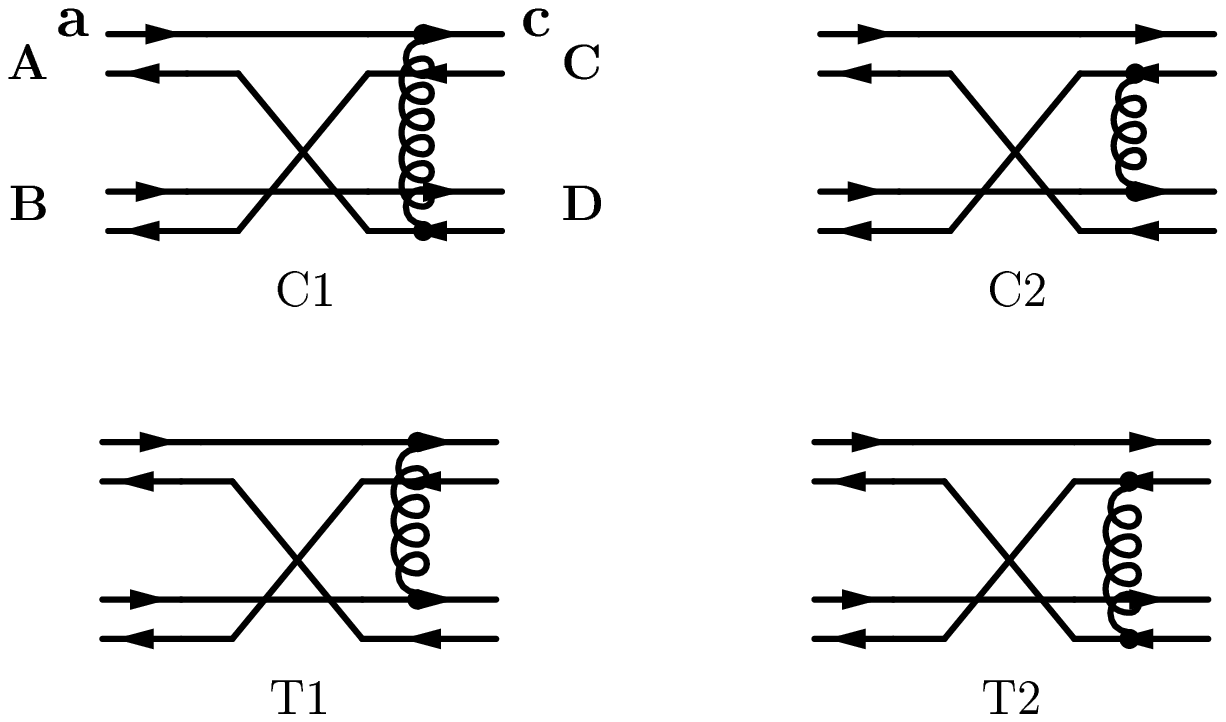}
\vspace*{0.4cm}\hspace*{2cm}
\begin{minipage}[t]{12cm}
\noindent {\bf Fig.\ 2}.  {`Post' diagrams for Born-order 
meson-meson scattering.}
\end{minipage}
\vskip 4truemm
\noindent 

One may prove in the context of non-relativistic quantum mechanics
that the `prior' and `post' diagrams give the same scattering
amplitude and hence the same cross section, provided that exact bound
state wavefunctions of the various $\{ H^{(0)}\} $ are used for the
external states \cite{Sch68}.  (This is discussed in detail and is
demonstrated numerically in Ref.\cite{Swa92} for $\pi\pi \to \rho\rho$
scattering.)  A consistent calculation thus leads to
description-identical results for the scattering amplitude in
non-relativistic quantum mechanics.  We shall confirm the
``prior-post'' equivalence numerically in our non-relativistic
calculations of the $J/\psi$ and $\psi'$ dissociation cross sections.

\section {Evaluation of the matrix element $I_{\rm {\lowercase {spin-space}}}$}

For the processes considered here, it suffices to treat reactions of
the form $A(12) + B(34) \to C(14) +D(32)$, in which constituents
(antiquarks) 2 and 4 are interchanged, as depicted in Figs. 1 and 2.
We denote the total angular momentum, the orbital angular momentum,
and the spin of meson $a$ ($a=A,B,C,$ and $D$) by $J_a$, $L_a$, and
$S_a$ respectively, with the associated spatial wavefunction $\Phi_a$
and spin wavefunction $\chi_a$.

The quantity $I_{\rm {\lowercase {spin-space}}}$ is the matrix element
of $V_{ij}(r_{ij})$ between the initial and final meson states.  The
interaction $V_{ij}(r_{ij})$ is the spin and spatial part of $H_{ij}$
(Eqs.\ (\ref{eq:Hij}) and (\ref{eq:vij})), and it consists of the sum
\begin{eqnarray}
\label{eq:vsum}
V_{ij}(r_{ij})=\sum_i^3 v_r^{(i)} v_s^{(i)}+V_{\rm con},
\end{eqnarray}
where the superscripts ${(i)}$ represent the color-Coulomb, linear,
and spin-spin interactions respectively.  Specifically,
$v_s^{(1)}=v_s^{(2)}=1$, $v_s^{(3)}=\bbox{s}_i\cdot \bbox{s}_j$, and
the corresponding $v_r^{(i)}$ can be determined from Eqs.\
(\ref{eq:Hij}) and (\ref{eq:vij}).  For the scattering problem the sum
of the amplitudes of all diagrams arising from the constant term
$V_{\rm con}$ is zero, so we do not need to include $V_{\rm con}$ in
deriving scattering amplitudes and matrix elements.

The matrix element $I_{\rm spin-space}$ is therefore the sum of three
terms, each of which is of the form
\begin{eqnarray}
\label{eq:vrvs}
\langle (\Phi_C\chi_C)^{J_C}_{J_{Cz}} (&\Phi_D&\chi_D)^{J_D}_{J_{Dz}} |   
v_r v_s |\, (\Phi_A\chi_A)^{J_A}_{J_{Az}}
(\Phi_B\chi_B)^{J_B}_{J_{Bz}} \rangle \nonumber\\
&=&\sum_{JJ_zJ'J_z'} 
(J_C\, J_{Cz}\, J_D\, J_{Dz}|J'\, J_z')
(J\, J_z\, |J_A\, J_{Az}\, J_B\, J_{Bz}) \nonumber\\
& &
~~~~~~~~\times
\langle \, \left [ (\Phi_C\chi_C)^{J_C} (\Phi_D\chi_D)^{J_D} \right
] _{J_z'}^{J'}| v_r v_s |\, 
\left [ (\Phi_A\chi_B)^{J_A} (\Phi_B\chi_B)^{J_B} \right
] _{J_z}^J \,  \rangle .
\end{eqnarray}   
In the above equation, the coupled initial state $| \left [
(\Phi_A\chi_B)^{J_A} (\Phi_B\chi_B)^{J_B} \right ]^J_{J_z}\rangle
\equiv |\Psi_{\rm in}^{JJ_z} \rangle $ of mesons $A(12)$ and $B(34)$
can be written as \cite{Tal63}
\begin{eqnarray}
|\Psi_{\rm in}^{JJ_z} \rangle
&=&\sum_{S,L}\langle (\chi_A \chi_B)^S (\Phi_A\Phi_B)^LJJ_z|
(\Phi_A\chi_B)^{J_A} (\Phi_B\chi_B)^{J_B}JJ_z \rangle
 |\, \left [ (\chi_A\chi_B)^S (\Phi_A\Phi_B)^L \right ] _{J_z}^J \, \rangle
\nonumber \\
&=& \sum_{SLS_z} 
\hat S \hat L \hat {J_A} \hat {J_B}
\left \{ 
\matrix{  S_A   &    S_B   &  S  \cr
          L_A   &    L_B   &  L  \cr   
          J_A   &    J_B   & J  \cr } 
\right \}
(S ~ S_z ~L ~(J_z-S_z)| J~J_z )~|\, (\chi_A\chi_B)^S_{S_z} (\Phi_A\Phi_B)
^L_{(J_z-S_z)} \,\rangle,
\end{eqnarray}
where ${\hat S} \equiv \sqrt{2S+1}$.  The final state $| \left [
(\Phi_C\chi_C)^{J_C} (\Phi_D\chi_D)^{J_D} \right ]_{J_z'}^{J'} \rangle
\equiv |\Psi_{\rm final}^{J'J_z'} \rangle $ of mesons $C(14)$ and
$D(32)$ can be written in a similar way, so the matrix element on the
righthand side of Eq.\ (\ref{eq:vrvs}) is
\begin{eqnarray}
\label{eq:fovlp}
\langle \Psi_{\rm final}^{J'J_z'} | v_r\, v_s  | \Psi_{\rm
in}^{JJ_z} \rangle &=& \sum_{SLS_zS'L'S_z'}
\hat S \hat L \hat {J_A} \hat {J_B} 
\left \{ \protect\matrix{ S_A &
S_B & S \cr L_A & L_B & L \cr J_A & J_B & J \cr} \right \} 
\hat {S'}  \hat {L'} \hat {J_C} \hat {J_D}
\left \{ \matrix{
S_C & S_D & S' \cr L_C & L_D & L' \cr J_C & J_D & J' \cr}\right \}
\nonumber \\
& \times & (S ~ S_z ~L ~(J_z-S_z)| J~J_z ) (S' ~ S_z' ~L'
~(J_z'-S_z')| J'~J_z' )^* \nonumber \\ & \times & \langle
(\Phi_C\Phi_D)^{L'}_{(J_z'-S_z')} |v_r| (\Phi_A\Phi_B) ^L_{(J_z-S_z)}
\rangle ~\langle (\chi_C\chi_D)^{S'}_{S_z'} |v_s|
(\chi_A\chi_B)^S_{S_z}\rangle.
\end{eqnarray}
The above result shows that $I_{\rm spin-space}$ is in general a sum
of products of a spatial matrix element $\langle
(\Phi_C\Phi_D)^{L'}_{(J_z'-S_z')} |v_r| (\Phi_A\Phi_B)^L_{(J_z-S_z)}
\rangle$ and a spin matrix element $I_{\rm spin}=\langle
(\chi_C\chi_D)^{S'}_{S_z'} |v_s| (\chi_A\chi_B) ^S_{S_z}\rangle$.  For
our interaction, the spin matrix element $\langle (\chi_C\chi_D)^{S'}
_{S_z'} |v_s| (\chi_A\chi_B)^S_{S_z}\rangle$ is diagonal in $S$ and
$S_z$, and is independent of $S_z$, as shown in the next section.

In this paper, we shall specialize to the cases in which mesons $B$,
$C$ and $D$ are all $S$-wave mesons with $L_B=L_C=L_D=0$.  Therefore
we have $L=L_A$, $J_B=S_B$, $J'=S$, $J_z'=S_z$, and 
\begin{eqnarray}
\label{eq:ovlp}
\langle \Psi_{\rm final}^{S S_z} | v_r v_s |
\Psi_{\rm in}^{JJ_z} \rangle
&=& 
\hat S \hat {L_A} \hat {J_A} \hat {J_B}
\left \{ \matrix{ S_A  &   S_B  &   S  \cr
       L_A  &   0  &   L_A  \cr
        J_A  &   S_B  &   J  \cr} \right \}
(S ~ S_z ~L_A ~(J_z-S_z)| J~J_z ) 
\nonumber \\& & \times
\langle 
\Phi_C\Phi_D |v_r| (\Phi_A\Phi_B)^{L_A}_{(J_z-S_z)} \rangle
\langle (\chi_C\chi_D)^{S}_{S_z} |v_s| (\chi_A\chi_B)^S_{S_z}\rangle,
\end{eqnarray}
where $|S_A-S_B| \le S \le (S_A+S_B)$ and $|S-L_A| \le J \le (S+L_A)$.
For the collision of unpolarized mesons, we can calculate the square
of the matrix element, $|I_{\rm space-spin}|^2$, average it over the
initial states, and sum it over the final states.  The result is
\begin{eqnarray}
&&\overline {|I_{\rm space-spin}|^2}
={1 \over (2J_A+1)(2S_B+1)}\sum_{JJ_zSS_z}(\hat S \hat L_A \hat {J_A}
\hat {S_B} )^2 
\left \{
\matrix{  S_A   &    S_B   &  S  \cr
          L_A   &    0  &  L_A  \cr
          J_A   &    S_B   & J  \cr }
\right \}^2  \nonumber\\
&&\times |(S ~ S_z ~L_A ~(J_z-S_z)| J~J_z )|^2
\left | 
\sum_i^3 
\langle \Phi_C\Phi_D |v_r^{(i)}| (\Phi_A\Phi_B)^{L_A}_{(J_z-S_z)} \rangle
\langle (\chi_C\chi_D)^{S}_{S_z} |v_s^{(i)}|(\chi_A\chi_B)^S_{S_z}\rangle
\right |^2.
\end{eqnarray}
The summation over $S_z$ can be carried out and the summation over
$J_z$ can be converted to a summation over $M_A$. We then obtain
\begin{eqnarray}
\overline {|I_{\rm space-spin}|^2}
&=&
\sum_{S\, J\, M_A} 
(\hat S \hat J)^2
\left \{
\matrix{  S_A   &    S_B   &  S  \cr
          L_A   &    0  &  L_A  \cr
          J_A   &    S_B   & J  \cr }
\right \}^2
\nonumber\\
& &~~~~~~~~~~~~~\times
\left |
\sum_i^3
\langle \Phi_C\Phi_D |v_r^{(i)}| (\Phi_A\Phi_B)^{L_A}_{M_A} \rangle
\langle (\chi_C\chi_D)^{S}_{S_z} |v_s^{(i)}|(\chi_A\chi_B)^S_{S_z}\rangle
\right |^2,                                                                    
\end{eqnarray}
where $-L_A \le M_A \le L_A$.  From the relation between the matrix
element of $V_{ij}(r_{ij})$ and the cross section, the above result
leads to the following ``unpolarized'' cross section for the
collision of unpolarized mesons,
\begin{eqnarray}
\label{eq:final}
\sigma^{\rm unpol}
=
\sum_{S\, J\, M_A}
(\hat S \hat J)^2
\left \{
\matrix{  S_A   &    S_B   &  S  \cr
          L_A   &    0  &  L_A  \cr
          J_A   &    S_B   & J  \cr }
\right \}^2  
\sigma(L_A M_A S S_z)
\end{eqnarray} 
where $\sigma(L_A M_A S S_z)$ is the cross section for the initial
meson system to have a total internal orbital angular momentum $L_A$
and total spin $S$ with azimuthal components $M_A$ and $S_z$
respectively.  For our interaction of Eq.\ (\ref{eq:Hij}), $\sigma(L_A
M_A S S_z)$ is independent of $S_z$, and thus the label $S_z$ can be
omitted.  We can write out the results for other simple unpolarized
cases.  If $L_A\ne 0$ and $S_B=0$, then $S=S_A$ and the result of Eq.\
(\ref{eq:final}) becomes
\begin{eqnarray}
\sigma^{\rm unpol} ={1 \over (2L_A+1)} \sum_{M_A=-L_A}^{L_A}
\sigma(L_A M_A S_A).
\end{eqnarray}                                                                 
If $L_A=0$ and $S_B \ne 0$, then the result of Eq.\ (\ref{eq:final}) is 
\begin{eqnarray}
\label{eq:unpola}
\sigma^{\rm unpol}
={1 \over (2S_A+1)(2S_B+1)}\sum_{S}(2S+1)
\sigma(S),
\end{eqnarray}
where $\sigma(S)$ is the cross section when the initial
two-meson system has a total spin $S$.

\section {Evaluation of the spin matrix element}

We denote the spins of the constituents in the scattering process
$A(12)B(34)\to C(14)D(23)$ by $s_1$, $s_2$, $s_3$, and $s_4$.
Using properties of the Wigner $\{9j\}$ symbols \cite{Tal63}, we may
rearrange the spins to obtain
\begin{eqnarray}
&&| (\chi_A\chi_B)^S_{S_z}\rangle
=|\left [(s_1s_2)S_A(s_3s_4)S_B \right ]^S_{S_z}\rangle
\nonumber\\
&=&(-1)^{S_B-s_4-s_3}\sum_{{S_{14}}~{S_{23}}} 
{\hat S}_A{\hat S}_B{\hat S}_{14}{\hat S}_{23}
\left \{ \matrix{ s_1  &   s_2  &   S_A  \cr
       s_4  &   s_3  &   S_B  \cr
        S_{14}  &   S_{23}  &   S \cr} \right \}
|[(s_1s_4)S_{14}(s_2s_3)S_{23}]^S_{S_z}\rangle
\end{eqnarray}
The phase factor
$(-1)^{S_B-s_4-s_3}$ arises from an interchange of spins in the
Clebsch-Gordon coefficients,
\begin{eqnarray}
|(s_3s_4)S_B \rangle =(-1)^{S_B-s_4-s_3}|(s_4s_3)S_B \rangle.
\end{eqnarray}
The matrix element of the spin unit operator $v_s=1$ 
is then given by
\begin{eqnarray}
\label{unit}
&&
\langle
(\chi_C\chi_D)^{S'}_{S_z'} |v_s| (\chi_A\chi_B)^S_{S_z}\rangle=
\langle 
 [(s_1s_4)S_C(s_3s_2)S_D]^{S'}_{S_z'} | 1 |
[(s_1s_2)S_A(s_3s_4)S_B]^S_{S_z} \rangle 
\nonumber \\
&&=\delta_{S S'} \delta_{S_z S_z'}
(-1)^{S_B+S_D-s_2-s_4-2s_3}
{\hat S}_A{\hat S}_B{\hat S}_C{\hat S}_D
\left \{ \matrix{ s_1  &   s_2  &   S_A  \cr
       s_4  &   s_3  &   S_B  \cr
        S_{C}  &   S_{D}  &   S \cr} \right \}.
\end{eqnarray}
The matrix element of the operator $v_s={\bf s}_i\cdot {\bf s}_j$ can
be derived similarly.  For diagrams C1 and C2, the matrix element is
given by
\begin{eqnarray}
\label{c1c2}
&&
\langle
(\chi_C\chi_D)^{S'}_{S_z'} |v_s| (\chi_A\chi_B)^S_{S_z}\rangle=
\langle 
 [(s_1s_4)S_C(s_3s_2)S_D]^{S'}_{S_z'}
 |{\bf s}_i\cdot {\bf s}_j |
[(s_1s_2)S_A(s_3s_4)S_B]^S_{S_z}
 \rangle
\nonumber \\
&&=\delta_{S S'} \delta_{S_z S_z'}
(-1)^{S_B+S_D-s_2-s_4-2s_3}
{\hat S}_A{\hat S}_B{\hat S}_C{\hat S}_D
\nonumber\\
&&\times
\left \{ \matrix{ s_1  &   s_2  &   S_A  \cr
       s_4  &   s_3  &   S_B  \cr
        S_{C}  &   S_{D}  &   S \cr} \right \}
{1\over 2} \left [ S_{ij}(S_{ij}+1)-S_i (S_i+1) -S_j (S_j+1) \right ].
\end{eqnarray}
The values of $i$, $j$, and $S_{ij}$ for diagrams C1 and C2 are
listed in Table I.
\vskip 0.5cm \centerline{Table I.  The values of $i$, $j$, and
$S_{ij}$ in Eq.\ (\ref{c1c2}).}  
{\vskip 0.4cm\hskip 5cm
\begin{tabular}{|c|c|c|c|} \hline
\label{tb11}
{\rm Diagram} 	& ~~$i$~~~  & ~~$j$~~~&  $S_{ij}$     \\
\hline
Prior C1&	1	&  	4	&	$S_C$	\\
Prior C2&  	2	&	3 	&	$S_D$	\\
\hline
Post C1 &	1	&	2 	& 	$S_A$	\\
Post C2	&	4	&	3 	&	$S_B$	\\
\hline
\end{tabular}
}
\vskip 0.6cm

The matrix element of $v_s={\bf s}_i\cdot {\bf s}_j$ for diagrams T1
and T2 is somewhat more complicated, and can be shown to be
\begin{eqnarray}
\label{ss}
&&\langle
(\chi_C\chi_D)^{S'}_{S_z'} |v_s| (\chi_A\chi_B)^S_{S_z}\rangle=
\langle 
 [(s_1s_4)S_C(s_3s_2)S_D]^S_{S_z} 
|{\bf s}_i\cdot {\bf s}_j  |
[(s_1s_2)S_A(s_3s_4)S_B]^{S'}_{S_z'} 
\rangle
\nonumber\\
&&=\delta_{S S'} \delta_{S_z S_z'}
\sum_{{S_{13}}~{S_{24}}}
(-1)^{S_{24}-s_4-s_2}
(2S_{13}+1) (2S_{24}+1){\hat S}_A{\hat S}_B{\hat S}_C{\hat S}_D
\nonumber \\
&&\times
\left \{ \matrix{ s_1  &   s_2  &   S_A  \cr
       s_3  &   s_4  &   S_B  \cr
        S_{13}  &   S_{24}  &   S \cr} \right \}
\left \{ \matrix{ s_1  &   s_4  &   S_C  \cr
       s_3  &   s_2  &   S_D  \cr
        S_{13}  &   S_{24}  &   S \cr} \right \}
{1\over 2} \left [ S_{ij}(S_{ij}+1)-S_i (S_i+1) -S_j (S_j+1) \right ],
\end{eqnarray}
where $ i$=1 and $j$=3 for diagram T1, and $i$=2 and $j$=4 for diagram
T2.  The quantities $S_{13}$ and $S_{24}$ span the full allowed range in
this summation.

Eqs.\ (\ref{unit}), (\ref{c1c2}), and (\ref{ss}) give the general
results for the spin matrix element $I_{\rm spin}$ of the unit
operator and the ${\bf s}_i\cdot {\bf s}_j$ operator in a
rearrangement collision.  Our results agree with the explicit
coefficients given in Table I of Barnes and Swanson \cite{Bar92}.
 
\section {Evaluation of the spatial matrix element}
\label{section:mod}

In the quark-interchange reaction of Eq.\ (\ref{eq:1}), the masses of
the quarks and antiquarks are different in general.  Previously, meson
scattering calculations with unequal masses using this approach had
been discussed in detail in coordinate space \cite{Swa92}.  Here we
give the corresponding momentum space results for general quark and
antiquark masses.

The spatial matrix element in    
Eq.\ (\ref{eq:fovlp}) is
\begin{eqnarray}
&& \langle (\Phi_C\Phi_D)^{L'}_{(J_z'-S_z')} |v_r| (\Phi_A\Phi_B)
^{L}_{(J_z-S_z)} \rangle \nonumber \\
&&=\sum_{M_A M_B M_C M_D}
(L_C M_C L_D  M_D |L'~(J_z'-S_z'))^*  
(L_A M_A L_B  M_B |L ~(J_z-S_z))
\nonumber \\
&&~~~~~~~~~~~~~~~~\times
 \langle \Phi_C(L_CM_C) \Phi_D(L_DM_D) |v_r|
\Phi_A(L_AM_A)\Phi_B(L_BM_B)) \rangle .
\end{eqnarray}
For the four diagrams in the reaction $A+B\rightarrow C+D$, the
spatial matrix element
\begin{eqnarray}
I_{\rm space}= \langle \Phi_C(L_CM_C) \Phi_D(L_DM_D)
|v_r| \Phi_A(L_AM_A)\Phi_B(L_BM_B))
\end{eqnarray}
can be written in the form
\begin{eqnarray}
\label{eq:eq1}
&&I_{space}
=\int\!\!\!\int d\bbox{\kappa} ~ d\bbox{\kappa}'
~\Phi_A [\zeta(2{\bbox{k}}_A-{\bbox{K}}_A)] 
\Phi_B [\zeta(2{\bbox{k}}_B-{\bbox{K}}_B)]\nonumber \\
&&~~~~~~~~\times\, \Phi_C [\zeta(2{\bbox{k}}_C-{\bbox{K}}_C)] 
\Phi_D [\zeta(2{\bbox{k}}_D-{\bbox{K}}_D)] 
V(\bbox{\kappa}' -\bbox{\kappa}).
\end{eqnarray}
Here the momentum arguments are shown explicitly, and the angular
momentum quantum numbers $L_i$ and $M_i$ for each meson are implicit.
The quantity $\zeta=\pm 1$ is an overall sign which depends on the
diagram (see Table II).  The quantity $V(\bbox{q})$, where
$\bbox{q}=\bbox{\kappa}-\bbox{\kappa}'$, is the Fourier transform of
the interaction $V_{ij}(r_{ij})$ [the spin and spatial part of
$H_{ij}(\bbox{r}_{ij})$ in Eq.\ (\ref{eq:vij})],
\begin{eqnarray}
\label{eq:vq}
V({\bbox{q}})= \int d{\bbox{r}}~ e^{-i {\bbox{q}}\cdot {\bbox{r}}_{ij} }~ V_{ij}(r_{ij}).
\end{eqnarray}
The momenta $\bbox{\kappa}$ is the initial three-momenta of the
scattered constituent that is initially in meson $A$ and
$\bbox{\kappa}'$ is its final three-momenta.  The variables
$\{{\bbox{k}}_i,~(i=A,B,C,D)\}$ are either $\bbox{\kappa}$ or
$\bbox{\kappa}'$ depending on the diagram, as specified in Table II.
We shall use the bold-faced symbols $\bbox{A},\bbox{B}, \bbox{C},$ and
$\bbox{D}$ to represent the momenta of $A$, $B$, $C$, and $D$
respectively.  For simplicity we shall treat the scattering problem in
the center-of-mass frame, so that $\bbox{B}$=$-\bbox{A}$ and
$\bbox{D}$=$-\bbox{C}$.  The quantity $\{{\bbox{K}}_i\}$ is a function
of $\bbox{A}$, $\bbox{C}$, and the mass parameter $f_i$, which is a
function of the quark and antiquark masses in meson $i$.  The function
${\bbox{K}}_i({\bbox{A}},{\bbox{C}},f_i)$ is tabulated for each
diagram in Table II.  For diagrams T1 and T2, the post and prior
variables are identical and so do not need to be tabulated separately.
\vskip 0.5cm \centerline{Table II. Diagram-dependent momentum arguments in 
post and prior formalisms.}
\vskip 0.4cm

\begin{tabular}{|c|c|c|c|c|c|c|c|c|c|} \hline
\label{tb2}
{\rm Diagram}&  ~~$\zeta$~~ 
                & ~~${\bbox{k}}_A$~~  & ~~${\bbox{K}}_A$~~
		& ~~${\bbox{k}}_B$~~  & ~~${\bbox{K}}_B$~~	
		& ~~${\bbox{k}}_C$~~  & ~~${\bbox{K}}_C$~~	
		& ~~${\bbox{k}}_D$~~  & ~~${\bbox{K}}_D$~~ 	  \\
\hline
Prior C1& ~1 
        &  $\bbox{\kappa}$	& $f_A {\bbox{A}}$  
	&  $\bbox{\kappa}$	& $f_B {\bbox{A}} + 2 {\bbox{C}}$
	&  $\bbox{\kappa}'$	& $f_C {\bbox{C}}$
	&  $\bbox{\kappa}$	& $f_D {\bbox{C}} + 2 {\bbox{A}}$ \\
Prior C2& -1
        &  $\bbox{\kappa}$	& $f_A {\bbox{A}}$  
	&  $\bbox{\kappa}$	& $f_B {\bbox{A}} - 2 {\bbox{C}}$
	&  $\bbox{\kappa}$	& $-f_C {\bbox{C}} + 2 {\bbox{A}}$
	&  $\bbox{\kappa}'$	& $-f_D {\bbox{C}}$             \\
\hline
Post C1	& ~1
	&  $\bbox{\kappa}$	& $f_A {\bbox{A}}$  
	&  $\bbox{\kappa}'$	& $f_B {\bbox{A}} + 2 {\bbox{C}}$
	&  $\bbox{\kappa}'$	& $f_C {\bbox{C}}$
	&  $\bbox{\kappa}'$	& $f_D {\bbox{C}} + 2 {\bbox{A}}$ \\
Post C2	& -1
	&  $\bbox{\kappa}$	& $f_A {\bbox{A}}$  
	&  $\bbox{\kappa}'$	& $f_B {\bbox{A}} - 2 {\bbox{C}}$
	&  $\bbox{\kappa}$	& $-f_C {\bbox{C}} + 2 {\bbox{A}}$
	&  $\bbox{\kappa}$	& $-f_D {\bbox{C}}$             \\
\hline
T1	& ~1
	&  $\bbox{\kappa}$	& $f_A {\bbox{A}}$  
	&  $\bbox{\kappa}'$	& $f_B {\bbox{A}} + 2 {\bbox{C}}$
	&  $\bbox{\kappa}'$	& $f_C {\bbox{C}}$
	&  $\bbox{\kappa}$	& $f_D {\bbox{C}} + 2 {\bbox{A}}$ \\
\hline
T2	& -1
	&  $\bbox{\kappa}$	& $f_A {\bbox{A}}$  
	&  $\bbox{\kappa}'$	& $f_B {\bbox{A}} - 2 {\bbox{C}}$
	&  $\bbox{\kappa}$	& $-f_C {\bbox{C}} + 2 {\bbox{A}}$
	&  $\bbox{\kappa}'$	& $-f_D {\bbox{C}}$             \\
\hline
\end{tabular}
\vskip 0.8cm

The mass parameter $f_i$ is unity for mesons with equal quark and
antiquark masses.    For unequal masses,
the $\{f_i\}$ are tabulated
in Table III in terms of 
\begin{eqnarray}
\Delta_i= { m(q)_{i}-m({\bar q})_i \over m(q)_i+ m({\bar q})_i}.
\end{eqnarray}
The \{ $f_i$\} are
the same in post and prior formalisms.

\vskip 0.5cm
\centerline{Table III.  The mass parameters $f_i$ for each diagram.}

\vskip 0.5cm \hspace*{3.2cm}
\begin{tabular}{|c|c|c|c|c|} \hline
{\rm Diagram}= 	& C1	 & C2	& T1	& T2  \\
\hline
$f_A =$ 	& $1+\Delta_A$	& $1-\Delta_A$& $1+\Delta_A$ & $1-\Delta_A$ \\
\hline
$f_B =$ 	& $1-\Delta_B$	& $1+\Delta_B$& $1-\Delta_B$ & $1+\Delta_B$ \\
\hline
$f_C=$ 		& $1+\Delta_C$	& $1+\Delta_C$& $1+\Delta_C$ & $1+\Delta_C$ \\
\hline
$f_D=$ 		& $1-\Delta_D$	& $1-\Delta_D$& $1-\Delta_D$ & $1-\Delta_D$ \\
\hline
\end{tabular}
\vskip 0.6cm 

To evaluate the spatial overlap integrals, we expand each meson
wavefunction $\Phi(2{\bf p})$ as a linear combination of
(nonorthogonal) Gaussian basis functions $\phi_n(2\bbox{p})$ of
different widths as
\begin{eqnarray}
\label{eq:phi1}
\Phi(2\bbox{p})=\sum_{n=1}^N a_n \phi_n(2\bbox{p}),
\end{eqnarray}
where 
\begin{eqnarray}
\label{eq:phi2}
\phi_n(2 \bbox{p})=
N_n (2p)^l \sqrt{ {4 \pi \over {(2l+1)!!}} }\, Y_{lm}(\hat {\bbox{p}})
\exp \{ - { (2 \bbox{p})^2 \over 8 n \beta^2 } \}.
\end{eqnarray}
This expansion makes the spatial integrals tractable.  We choose to
normalize the basis function $\phi_n(2 \bbox{p})$ according to
\begin{eqnarray}
\label{eq:phinor}
\int d\bbox{p}\, |\phi_n ({\bf2\bbox{p}})|^2 = 1,
\end{eqnarray}
which leads to the normalization constant
\begin{eqnarray}
N_n= \left ( {1 \over \pi n \beta^2} \right ) ^ {3/4} {1 \over (2 n
\beta^2)^{l/2}}.  
\end{eqnarray} 
We also normalize the meson wavefunction $\Phi(2\bbox{p})$ according to
\begin{eqnarray}
\label{eq:Phinor}
\int d\bbox{p}\, |\Phi ({\bf2\bbox{p}})|^2 = 1,
\end{eqnarray}
which implies a constraint on the coefficients $\{a_n\}$.

We shall first present our results for the spatial matrix element 
Eq.\ (\ref{eq:eq1}) for the case of all $S$-wave mesons, each
with a single Gaussian wavefunction of the type of Eq.\ (\ref{eq:phi2}),
\begin{eqnarray}
\label{eq:Phi}
\phi_i [\zeta(2{\bbox{k}}_i-{\bbox{K}}_i)]
=N_i\exp\{ - {\lambda_i \over 2}(2{\bbox{k}}_i-{\bbox{K}}_i)^2 \}
\end{eqnarray}
where $\lambda_i=1/4n\beta^2$ and
\begin{eqnarray}
N_i=
\sqrt{8} \left ( \lambda_i \over \pi \right )
^{3/4}.
\end{eqnarray}

The product of wavefunctions in Eq.\ (\ref{eq:eq1}) is explicitly
\begin{eqnarray}
& &
\phi_A [\zeta(2{\bbox{k}}_A-{\bbox{K}}_A)]\; 
\phi_B [\zeta(2{\bbox{k}}_B-{\bbox{K}}_B)]\;
\phi_C [\zeta(2{\bbox{k}}_C-{\bbox{K}}_C)]\;
\phi_D [\zeta(2{\bbox{k}}_D-{\bbox{K}}_D)]\;
\nonumber\\
& & ~~~~~~=N_A N_B N_C N_D 
\exp\{-\sum_{i=1}^4{\lambda_i \over 2}(2{\bbox{k}}_i-{\bbox{K}}_i)^2 \}.
\end{eqnarray}
The argument of the exponential, from the product of the four meson
wavefunctions, is a function of ${\bbox{k}}_i=\{ \bbox{\kappa} $,
$\bbox{\kappa}'\}$ and the quantities $\{ {\bbox{K}}_i \}$.  It can
also be written as a function of $\bbox{p}= ( \bbox{\kappa}+
\bbox{\kappa}')/2$ and ${\bbox{q}}= \bbox{\kappa}' - \bbox{\kappa}$.
In terms of $\bbox{p}$ and $\bbox{q}$, the $\{ {\bbox{k}}_i\} $ are
given by
\begin{eqnarray}
\label{eq:yi}
{\bbox{k}}_i=\bbox{p}-\eta_i~ {\bbox{q}}/2,
\end{eqnarray}
where $\eta_i$ is 
\begin{eqnarray}
\eta_i= \cases{ +1, & if~~ ${\bbox{k}}_i=\bbox{\kappa}\; $; \cr
             -1, & if~~ ${\bbox{k}}_i=\bbox{\kappa}'\; $. \cr}
\end{eqnarray}
Using Eq.\ (\ref{eq:yi}) 
and completing the square in the exponential, we obtain
\begin{eqnarray}
\label{eq:pro0}
\sum_{i=1}^4\; {\lambda_i \over 2}\; (2{\bbox{k}}_i-{\bbox{K}}_i)^2
=2\sum_i^4 \lambda_i\; ( \bbox{p} - \bbox{p}_0)^2 
     + {\mu\over 2}\;  ({\bbox{q}}-{\bbox{q}}_0)^2 + {\nu \over 2} ,
\end{eqnarray}
where the quantities $\bbox{p}_0$,
${\bbox{q}}_0$, $\mu$, and $\nu$ are defined below.
The shift $\bbox{p}_0$ is
\begin{eqnarray}
\label{eq:p0}
\bbox{p}_0=r_0 {\bbox{q}} + {\bf s}_0,
\end{eqnarray}
where $r_0$ and ${\bf s}_0$ are 
\begin{eqnarray}
r_0 = { \sum_{i=1}^4 \eta_i \lambda_i \biggl / 2 \sum_{i=1}^4 \lambda_i }\; ,
\end{eqnarray}
\begin{eqnarray}
{\bf s}_0 = {\sum_{i=1}^4 \eta_i {\bbox{K}}_i \biggl / 2 \sum_{i=1}^4 \lambda_i }\; .
\end{eqnarray}
The quantity $\mu$ in Eq.\ (\ref{eq:pro0}) is
\begin{eqnarray}
\mu = 4 \left ( \sum_{i=1}^4 {1 + \eta_i \over 2} \lambda_i \right )
 \left ( \sum_{j=1}^4 {1 - \eta_j \over 2} \lambda_j \biggl /
\sum_{i=1}^4 \lambda_i \right ),
\end{eqnarray}
the shift ${\bbox{q}}_0$ is
\begin{eqnarray}
{\bbox{q}}_0 = -{2 \over \mu \sum_{i=1}^4 \lambda_i}
\left [ \left (
\sum_{i=1}^4 {1 - \eta_i \over 2} \lambda_i \right )
\left ( \sum_{j=1}^4 {1 + \eta_j \over 2} \lambda_j {\bbox{K}}_j \right )
-
\left ( \sum_{i=1}^4 {1 + \eta_i \over 2} \lambda_i \right )
\left ( \sum_{j=1}^4 {1 - \eta_j \over 2} \lambda_j {\bbox{K}}_j \right )
\right ],
\end{eqnarray}
and $\nu$ is
\begin{eqnarray}
\nu=\sum_{i=1}^4 \lambda_i {\bbox{K}}_i^2
- 4 \sum_{i=1}^4 \lambda_i {\bf s}_0^2 - \mu {\bbox{q}}_0^2.
\end{eqnarray}
The product of wavefunctions in Eq.\ (\ref{eq:eq1}) can therefore be
written in a shifted Gaussian form,
\begin{eqnarray}
\label{eq:pro}
& &
~\phi_A [\zeta(2{\bbox{k}}_A-{\bbox{K}}_A)]\; 
\phi_B [\zeta(2{\bbox{k}}_B-{\bbox{K}}_B)]\;
\phi_C [\zeta(2{\bbox{k}}_C-{\bbox{K}}_C)]\; 
\phi_D [\zeta(2{\bbox{k}}_D-{\bbox{K}}_D)] 
\nonumber\\
& & ~~~~~~=N_A N_B N_C N_D 
\exp\{-2\sum_{i=1}^4 \lambda_i( \bbox{p} - \bbox{p}_0)^2 
     - {\mu\over 2} ({\bbox{q}}-{\bbox{q}}_0)^2 - {\nu \over 2} \}.
\end{eqnarray}
The spatial matrix element of Eq.\ (\ref{eq:eq1}) then becomes
\begin{eqnarray}
\label{eq:integ}
\int\!\!\!\int 
d\bbox{\kappa} ~ d\bbox{\kappa}'
~ \phi_A [\zeta(2{\bbox{k}}_A-{\bbox{K}}_A)] 
~ \phi_B [\zeta(2{\bbox{k}}_B-{\bbox{K}}_B)]
~ \phi_C [\zeta(2{\bbox{k}}_C-{\bbox{K}}_C)] 
~ \phi_D [\zeta(2{\bbox{k}}_D-{\bbox{K}}_D)] 
~V({\bbox{q}})
\nonumber \\
=
\int\!\!\!\int 
d\bbox{p} ~d{\bbox{q}} ~N_A N_B N_C N_D 
\exp\{-2\sum_{i=1}^4 \lambda_i( \bbox{p} - \bbox{p}_0)^2 
     - {\mu\over 2} ({\bbox{q}}-{\bbox{q}}_0)^2 - {\nu \over 2} \}\; V({\bbox{q}}).
\end{eqnarray}
The integration over $\bbox{p}$ can be carried out analytically, which gives
\begin{eqnarray}
\int\!\!\!\int 
d\bbox{p} ~d{\bbox{q}}~\prod_{i=1}^4
\phi_i [\zeta(2{\bbox{k}}_i-{\bbox{K}}_i)] \; V({\bbox{q}})
=
N_A N_B N_C N_D 
\left ( \pi \over 2 \sum_{i=1}^4 \lambda_i^4 \right)^{3/2} e^{- {\nu \over 2}}
\int d{\bbox{q}}~ e^{-{\mu \over 2} ( {\bbox{q}}-{\bbox{q}}_0)^2}\; V({\bbox{q}}).
\end{eqnarray}
Thus, the six-dimensional integral of Eq.\ (\ref{eq:integ})
is simplified to a three-dimensional integral involving an
integration over $V(\bbox{q})$.

The interaction $V({\bbox{q}})$, which we take from the standard quark
model $V_{ij}(r)$, is the sum of Fourier transforms of color-Coulomb,
spin-spin contact, linear confinement and constant terms.  The sum of
all diagrams arising from the constant term is zero for the scattering
problem, so we do not need to include $V_{\rm con}$ in deriving
scattering amplitudes.

For the remaining three interactions, we have (using some integrals
of Ref.\cite{Bar99})
\begin{eqnarray}
\int d{\bbox{q}} ~ e^{-{\mu \over 2}\; ( {\bbox{q}}-{\bbox{q}}_0)^2 } V_{\rm
Cou.}({\bbox{q}})
&=&\int d{\bbox{q}}~ e^{-{\mu \over 2} ({\bbox{q}}-{\bbox{q}}_0)^2 }{4 \pi
\alpha_s \over {\bbox{q}}^2}
\nonumber\\
&=& {4 \pi \alpha_s (2 \pi)^{3/2} \over \sqrt{\mu}}\;
e^{-\mu {\bbox{q}}_0^2/2}  {}_1F_1 ( {1\over 2}, {3\over 2}, {\mu {\bf
q}_0^2 \over 2} ), 
\end{eqnarray}
\begin{eqnarray}
\int d{\bbox{q}}\; e^{-{\mu \over 2} ( {\bbox{q}}-{\bbox{q}}_0)^2}
\; V_{\rm ss}({\bf
q})
&=&\int d{\bbox{q}}\;  e^{-{\mu \over 2} ( {\bbox{q}}-{\bbox{q}}_0)^2}
\left (
{-8 \pi \alpha_s \over 3 m_i m_j} \right )
e^{-{\bbox{q}}^2 / 4 \sigma^2 }
\nonumber\\
&=& {-8 \pi \alpha_s \over 3 m_i m_j} 
\left ( { 2 \pi \over \mu } \right ) ^{3/2} \left ( {2 \sigma^2 \mu
\over 1 + 2 \sigma^2 \mu } \right )^{3/2} 
\exp \{- {\mu {\bbox{q}}_0^2 \over 2(1 + 2 \sigma^2 \mu)} \} ,
\end{eqnarray}
and
\begin{eqnarray}
\int d{\bbox{q}}~ e^{-{\mu \over 2} ( {\bbox{q}}-{\bbox{q}}_0)^2}~ V_{\rm lin}({\bf
q})
=\left ( -{ 3\over 4 }\right )  8 \pi b \left ( {2 \pi \over 3} \right
)^{3/2} \mu^2 e^{-\mu {\bbox{q}}_o^2/2} {}_1F_1 \left ( -{1\over 2}, {3\over 2}, {\mu
{\bbox{q}}_0^2 \over 2} \right ),
\end{eqnarray}
where $V_{\rm lin}({\bbox{q}})=(-3/4)\int d{\bf r}\,
e^{-i{\bbox{q}}\cdot {\bf r}}\, b r $.  These results allow one to
evaluate the transition matrix elements $T_{fi}$ explicitly for the
different interactions in diagrams C1, C2, T1, and T2 for the case of
Gaussian meson wavefunctions.

The wavefunctions we employ here are in general sums of Gaussians of
different widths (Eq.\ (\ref{eq:phi1})).  Eq.\ (\ref{eq:eq1}) can be
evaluated in that case as well, so that the spatial matrix element
Eq.\ (\ref{eq:eq1}) becomes
\begin{eqnarray}
&&I_{space}\nonumber\\
&&=
\int 
\!\!\!
\int
d\bbox{\kappa}\; d\bbox{\kappa}'\;
\Phi_A [\zeta(2{\bbox{k}}_A-{\bbox{K}}_A)]\; 
\Phi_B [\zeta(2{\bbox{k}}_B-{\bbox{K}}_B)]\;
\Phi_C [\zeta(2{\bbox{k}}_C-{\bbox{K}}_C)]\; 
\Phi_D [\zeta(2{\bbox{k}}_D-{\bbox{K}}_D)]\; 
V(\bbox{\kappa}'-\bbox{\kappa})
\nonumber\\
&&=\sum_{n_A=1}^N \sum_{n_B=1}^N \sum_{n_C=1}^N \sum_{n_D=1}^N
a_{n_A}a_{n_B}a_{n_C}a_{n_D}\; I_{space}(n_A,n_B,n_C,n_D),
\end{eqnarray}
where 
\begin{eqnarray}
I_{space}(n_A,n_B,n_C,n_D)
&=&
\int 
\!\!\!
\int
d\bbox{\kappa}\; d\bbox{\kappa}'\;
\phi_{An_A} [\zeta(2{\bbox{k}}_A-{\bbox{K}}_A)]\; 
\phi_{Bn_B} [\zeta(2{\bbox{k}}_B-{\bbox{K}}_B)]
\nonumber \\
& &\times ~\phi_{Cn_C} [\zeta(2{\bbox{k}}_C-{\bbox{K}}_C)]\; 
\phi_{Dn_D} [\zeta(2{\bbox{k}}_D-{\bbox{K}}_D)]\; 
V(\bbox{\kappa}'-\bbox{\kappa}).
\end{eqnarray}
$I_{space}(n_A,n_B,n_C,n_C)$ is the previous result of Eq.\
(\ref{eq:eq1}) for a single component wavefunction. The overlap
integral in the multicomponent case is simply a sum of
single-component contributions, each weighed by a coefficient product
$a_{n_A} a_{n_B} a_{n_C} a_{n_D}$.

After the matrix elements for the interaction (\ref{eq:Hij}) are
evaluated, the cross section for the process $A+B\rightarrow C+D$ can
then be obtained using conventional scattering theory, as discussed in
Section II.

\section {Evaluation of the spatial matrix element 
           for an $\bbox{L}=1$ Meson}

In the last section, we considered the scattering of $S$-wave ($L=0$)
mesons.  Here we generalize to collisions in which a $P$-wave ($L=1$)
meson $A$ collides with an $S$-wave meson $B$, and scatter into two
$S$-wave mesons $C$ and $D$.

First we consider single-component Gaussian wavefunctions.  (The
results can be easily generalized to multicomponent Gaussian
wavefunctions.) Eq.\ (\ref{eq:eq1}) becomes
\begin{eqnarray}
&&\int  
\!\!\!
\int
d\bbox{\kappa} ~ d\bbox{\kappa}'\;
\phi_A (2{\bbox{k}}_A-{\bbox{K}}_A)\; 
\phi_B(2{\bbox{k}}_B-{\bbox{K}}_B)\;
\phi_C(2{\bbox{k}}_C-{\bbox{K}}_C)\; 
\phi_D (2{\bbox{k}}_D-{\bbox{K}}_D)\; V({\bbox{q}})
\nonumber \\
&&= 
\int 
\!\!\!
\int
d\bbox{\kappa} ~ d\bbox{\kappa}'
N_A\,
|2\bbox{p}_A|^{L_A} \sqrt{ {4 \pi \over {(2L_A+1)!!}} }\;
Y_{L_A M_A}(\hat {\bbox{p}}_A)
\exp\{ - {\lambda_i \over 2}(2{\bbox{k}}_A-{\bbox{K}}_A)^2 \}
\nonumber \\
&& \times
\phi_B(2{\bbox{k}}_B-{\bbox{K}}_B)\;
\phi_C(2{\bbox{k}}_C-{\bbox{K}}_C)\; 
\phi_D (2{\bbox{k}}_D-{\bbox{K}}_D)\;
 V({\bbox{q}}),
\nonumber 
\end{eqnarray}
where $2\bbox{p}_A=2{\bbox{k}}_A-{\bbox{K}}_A$.
Setting $L_A=1$ for the $P$-wave meson $A$, we have
\begin{eqnarray}
|2{\bbox{k}}_A-{\bbox{K}}_A|^{L_A}\, \sqrt{ {4 \pi \over {(2L_A+1)!!}} }\;
Y_{L_AM_A}(\hat {\bbox{p}}_A)
= \cases { 2\kappa_z - f_A A_z & {\rm~ if ~$M_A$=~0}\cr
         -(2\kappa_x - f_A A_x)-i(2\kappa_y - f_A A_y) & {\rm~ if ~$M_A$=~1}\cr
          (2\kappa_x - f_A A_x)-i(2\kappa_y - f_A A_y) & {\rm~ if ~$M_A$=-1}\cr}.
\end{eqnarray}
It then suffices to evaluate 
\begin{eqnarray} 
& &\int 
\!\!\!
\int
d\bbox{\kappa} ~ d\bbox{\kappa}'
 (2\bbox{\kappa}-f_A {\bbox{A}})_k N_A
\exp\{ - {\lambda_A \over 2}(2{\bbox{k}}_A-{\bbox{K}}_A)^2 \}\; \hskip 6cm 
\nonumber\\
& &~~~~~~~~~\times 
\phi_B(2{\bbox{k}}_B-{\bbox{K}}_B)\;
\phi_C(2{\bbox{k}}_C-{\bbox{K}}_C)\; 
\phi_D (2{\bbox{k}}_D-{\bbox{K}}_D)\; V({\bbox{q}})
\nonumber \\
&=&
\int 
\!\!\!
\int
d\bbox{p}\; d\bbox{q}\;  
 (2\bbox{\kappa}-f_A {\bbox{A}})_k ~N_A N_B N_C N_D 
\exp\{-2\sum_{i=1}^4 \lambda_i( \bbox{p} - \bbox{p}_0)^2 
     - {\mu\over 2} 
 ({\bbox{q}}-{\bbox{q}}_0)^2 - {\nu \over 2} \}\; V({\bbox{q}}).
\end{eqnarray} 
We can express $2\bbox{\kappa}-f_A {\bbox{A}}$ in terms of $\bbox{p}$
and $\bbox{q}$;
\begin{eqnarray}
2\bbox{\kappa}-f_A {\bbox{A}}
=2\bbox{p}-{\bbox{q}} - f_A {\bbox{A}}
=2(\bbox{p}-\bbox{p}_0)+2\bbox{p}_0 -{\bbox{q}} - f_A {\bbox{A}}\ .
\end{eqnarray}
Substituting Eq.\ (\ref{eq:p0}) into this result,  we find 
\begin{eqnarray}
2\bbox{\kappa}-f_A {\bbox{A}}
=2(\bbox{p}-\bbox{p}_0)+(2r_0-1)({\bbox{q}}-{\bbox{q}}_0) 
+(2r_0-1){\bbox{q}}_0 +2{\bf s}_0  - f_A {\bbox{A}}\ .
\end{eqnarray}
The integral of $\bbox{p}-\bbox{p}_0$ gives zero.  The integration
over the last three terms can be carried out in the same way as in the
$L_A=0$ case, because ${\bbox{q}}_0$, ${\bf s}_0$ and $\bbox{A}$ are
independent of the integration variables.  It is thus only necessary
to evaluate the integral
\begin{eqnarray}
\int 
\!\!\!
\int
d\bbox{p}\; d\bbox{q}\; N_A N_B N_C N_D 
 ({\bbox{q}}-{\bbox{q}}_0)_k 
\exp\{-2\sum_{i=1}^4 \lambda_i( \bbox{p} - \bbox{p}_0)^2 
     - {\mu\over 2} ({\bbox{q}}-{\bbox{q}}_0)^2 - {\nu \over 2} \}\; V({\bbox{q}}).
\end{eqnarray}
The integration over $\bbox{p}$ can be carried out analytically, which
gives
\begin{eqnarray}
&&
\int 
\!\!\!
\int
d\bbox{p} ~d{\bbox{q}} ~N_A N_B N_C N_D 
 ({\bbox{q}}-{\bbox{q}}_0)_k 
\exp\{-2\sum_{i=1}^4 \lambda_i( \bbox{p} - \bbox{p}_0)^2 
     - {\mu\over 2} ({\bbox{q}}-{\bbox{q}}_0)^2 - {\nu \over 2} \}\; V({\bbox{q}})
\nonumber\\
=&&
N_A N_B N_C N_D 
\left ( \pi \over 2 \sum_{i=1}^4 \lambda_i^4 \right)^{3/2} e^{- {\nu \over 2}}
{1 \over \mu} {\partial \over \partial q_{0k}} 
\int d{\bbox{q}}~ 
e^{-{\mu \over 2} ( {\bbox{q}}-{\bbox{q}}_0)^2} V({\bbox{q}}).
\end{eqnarray}
The $d{\bbox{q}}$ integrals for the various potentials have already
been obtained in closed form, and the differentiation with respect to
$q_{0k}$ is straightforward.  We then find
\begin{eqnarray}
{\partial \over\partial q_{0k}} 
&&\int d{\bbox{q}}~ e^{-{\mu \over 2} ( {\bbox{q}}-{\bbox{q}}_0)^2 } 
\; 
V_{\rm
Cou.}({\bbox{q}})
\nonumber\\
&&= {4 \pi \alpha_s (2 \pi)^{3/2} \over \sqrt{\mu}}
\; 
e^{-\mu {\bbox{q}}_0^2/2}  
\left [-{}_1F_1 ( {1\over 2}, {3\over 2}, {\mu {\bbox{q}}_0^2 \over 2} ) 
+{1\over 3} {}_1F_1 ( {3\over 2}, {5\over 2}, {\mu {\bbox{q}}_0^2 \over
2} ) \right ] 
\mu\,  q_{0k},
\end{eqnarray}
\begin{eqnarray}
{\partial \over \partial q_{0k}}
\int d{\bbox{q}}~ e^{-{\mu \over 2} ( {\bbox{q}}-{\bbox{q}}_0)^2} V_{\rm ss}({\bf
q})
&&= {-8 \pi \alpha_s \over 3 m_i m_j} 
\left ( { 2 \pi \over \mu } \right ) ^{3/2} \left ( {2 \sigma^2 \mu
\over 1 + 2 \sigma^2 \mu } \right )^{3/2} 
\exp \{- {\mu {\bbox{q}}_0^2 \over 2(1 + 2 \sigma^2 \mu)} \} 
\nonumber \\
&&\times
\left ( - {\mu q_{oi} \over 1 + 2 \sigma^2 \mu} \right ),
\end{eqnarray}
and
\begin{eqnarray}
{\partial \over \partial q_{0k}}
\int d{\bbox{q}} ~e^{-{\mu \over 2} ( {\bbox{q}}-{\bbox{q}}_0)^2} V_{\rm lin}({\bf
q})
&=&\left ( -{ 3\over 4 }\right )b~  8 \pi \left ( {2 \pi \over 3} \right
)^{3/2} \mu^2 e^{-\mu {\bbox{q}}_o^2/2} 
\nonumber \\
& &\times
\left [-{}_1F_1 \left ( -{1\over 2}, {3\over 2}, {\mu {\bbox{q}}_0^2 \over 2}
\right )
-{1\over 3}{}_1F_1 \left ( {1\over 2}, {5\over 2}, {\mu {\bbox{q}}_0^2 \over 2}
\right ) \right ]\mu \, q_{0k}.
\end{eqnarray}
The scattering amplitude $T_{fi}$ and cross section $\sigma_{fi}$ for
the dissociation of a $P$-wave meson through an $SP\to SS$ reaction
will subsequently be evaluated using these results.

\section{Meson Wave Functions}
\label{section:bound}

In nonrelativistic reaction theory, the equality of the scattering
amplitude for rearrangement reactions in post and prior formalisms
follows if and only if the two-body interaction used in evaluating the
scattering matrix elements is identical to the interaction that
generates the bound state wavefunctions.  If this is not the case, the
post and prior scattering amplitudes will differ.  It is therefore
especially important to determine accurate bound state wavefunctions
in evaluating scattering amplitudes.  For this purpose, we will first
search for a set of interaction Hamiltonian parameters that fit the
known meson spectrum with reasonable accuracy. This interaction and
the associated bound state wavefunctions will be used in our
subsequent reaction calculations.

For a bound state of quark $i$ and antiquark $j$ of momentum 
$\bbox{p}_i$ 
and 
$\bbox{p}_j$ 
and reduced momentum $\bbox{p}$,
\begin{eqnarray}
\bbox{p} = { m_j \bbox{p}_i - m_i \bbox{p}_j \over m_i + m_j },
\end{eqnarray}
the Hamiltonian is 
\begin{eqnarray}
\label{eq:ham}
H={ \bbox{p}^2 \over 2 \mu} + V(r) ,
\end{eqnarray}
where $\mu$ is the reduced mass $m_i m_j /(m_i+m_j)$, and $ V(r)$
is the quark-antiquark interaction, (see $H_{ij}$
of Eq.\ (\ref{eq:Hij})),
\begin{eqnarray}
V(r)
=-{\bbox{\lambda}(i) \over 2}\cdot {\bbox{\lambda}^T(j) \over 2} \left \{
{\alpha_s \over r} - {3 b \over 4} r - {8 \pi \alpha_s \over
3 m_i m_j } \bbox{s}_i \cdot \bbox{s}_j \left ( {\sigma^3 \over
\pi^{3/2} } \right ) e^{-\sigma^2 r^2}  
+ V_{\rm con}
\right
\}.
\end{eqnarray}
For a quark and antiquark in a color-singlet hadron, 
the matrix element of
$-{\bbox{\lambda}}(i)\cdot{\bbox{\lambda}}^T(j) / 4$ is 
the familiar color factor $C_f=-4/3$.

For given orbital angular momentum quantum numbers $l$ and $m$, the
eigenstate $\Phi(2\bbox{p})$ of this Hamiltonian can be represented as
the expansion in Eq.\ (\ref{eq:phi1}) in the set of (nonorthogonal)
Gaussian basis states $\{\phi_n\}$ of Eq.\ (\ref{eq:phi2}) with
expansion coefficients $\{a_n\}$.  The eigenvalue equation
$H\Phi=E\Phi$ then becomes the matrix equation
\begin{eqnarray}
\label{eig}
{\cal H}a=E B a,
\end{eqnarray}
where $a$ is a column matrix with elements $\{a_1, a_2, ..., a_N
\}$, $B$ is the symmetric matrix 
\begin{equation}
B_{ij} = \langle i | j \rangle \equiv 
\int {d \bbox{p} }~ \phi_i^* (2\bbox{p})\, \phi_j (2\bbox{p})
=   \left ( {2 \sqrt{ij}\over i+j }  \right ) ^{l+3/2} ,
\end{equation}
and $\cal H$ is the Hamiltonian matrix 
\begin{eqnarray}
{\cal H}_{ij}
=T_{ij}+V_{ij},
\end{eqnarray}
which is the sum of the kinetic energy matrix $T_{ij}$ 
\begin{eqnarray}
T_{ij}=\int {d \bbox{p}}~ \phi_i^*(2\bbox{p})
~{ \bbox{p}^2 \over 2\mu}~ \phi_j (2\bbox{p})
= (2l+3)
{ij \over i+j}
 \left ( {2 \sqrt{ij}\over i+j }  \right ) ^{l+3/2}
{ \beta^2 \over 2 \mu} \ ,
\end{eqnarray}
and the potential matrix $V_{ij}$ 
\begin{eqnarray}
V_{ij} \equiv \langle i | V|j\rangle 
= (2\pi)^3 \int {d \bbox{r} }~  \tilde \phi_i^*(\bbox{r})~ V(\bbox{r})~ 
\tilde \phi_j(\bbox{r}),
\end{eqnarray}
where $\tilde \phi_i(\bbox{r})$ is the Fourier transform of
$\phi_i(2\bbox{p})$,
\begin{eqnarray}
\tilde \phi_i(\bbox{r})=\int {d \bbox{p} \over (2\pi)^3} 
\, e^{i \bbox{p} \cdot \bbox{r}}\,
\phi_i(2\bbox{p}).
\end{eqnarray}
Evaluation of the potential matrix elements for the color-Coulomb
interaction gives
\begin{eqnarray}
V_{ij} ({\rm Cou.})&=&
C_f 
4 \pi \alpha_s 
\langle i |~ 
{ 1\over r}
~| j \rangle 
\nonumber \\
&=& 
C_f 
{4 \pi \alpha_s  \beta 
\over
(2\pi)^{3/2} }
{ 2^l ~ l!\over (2l+1)!! }
\sqrt{i+j} \left ( {2 \sqrt{ij}\over i+j }  \right ) ^{l+3/2}.
\end{eqnarray}
For the linear interaction we similarly find
\begin{eqnarray}
V_{ij}({\rm lin.})&=&\langle i | ~{ C_f\left (- {3 \over 4} \right ) 
br }~ | j \rangle  \nonumber\\
&=&C_f\left (- {3 \over 4} \right )  
{b \over \beta }\;
{ 8 \pi  
\over 
(2\pi)^{3/2} 
}
{  (l+1)!
\over 
(2l+1)!! 
}
{1
\over 
\sqrt{i+j} }
\left ( {2 \sqrt{ij}\over i+j }  \right ) ^{l+3/2} \ ,
\end{eqnarray}
and for the spin-spin interaction we find
\begin{eqnarray}
V_{ij}({\rm ss})
&=&
 - C_f {8 \pi \alpha_s \over
3 m_i m_j } 
 \left ( {\sigma^3 \over
\pi^{3/2} } \right ) 
\langle i | ~
\bbox{s}_i \cdot \bbox{s}_j 
e^{-\sigma^2 r^2} 
~| j \rangle 
\nonumber \\
&=&-C_f {8 \pi \alpha_s \over
3 m_i m_j } \bbox{s}_i \cdot \bbox{s}_j 
\left ( {\sigma^3 \over
\pi^{3/2} } \right )
 \left ( {2 \sqrt{ij}\over i+j+2\sigma^2/\beta^2 }
  \right ) ^{l+3/2}.
\end{eqnarray}
Given these Hamiltonian matrix elements, the eigenvalues and
eigenvectors can be obtained from the eigenvalue equation (\ref{eig}).
In our numerical calculations we used a six-component ($N=6$) space of
Gaussian basis functions.

\vspace*{2.5cm}
\epsfxsize=300pt
\includegraphics{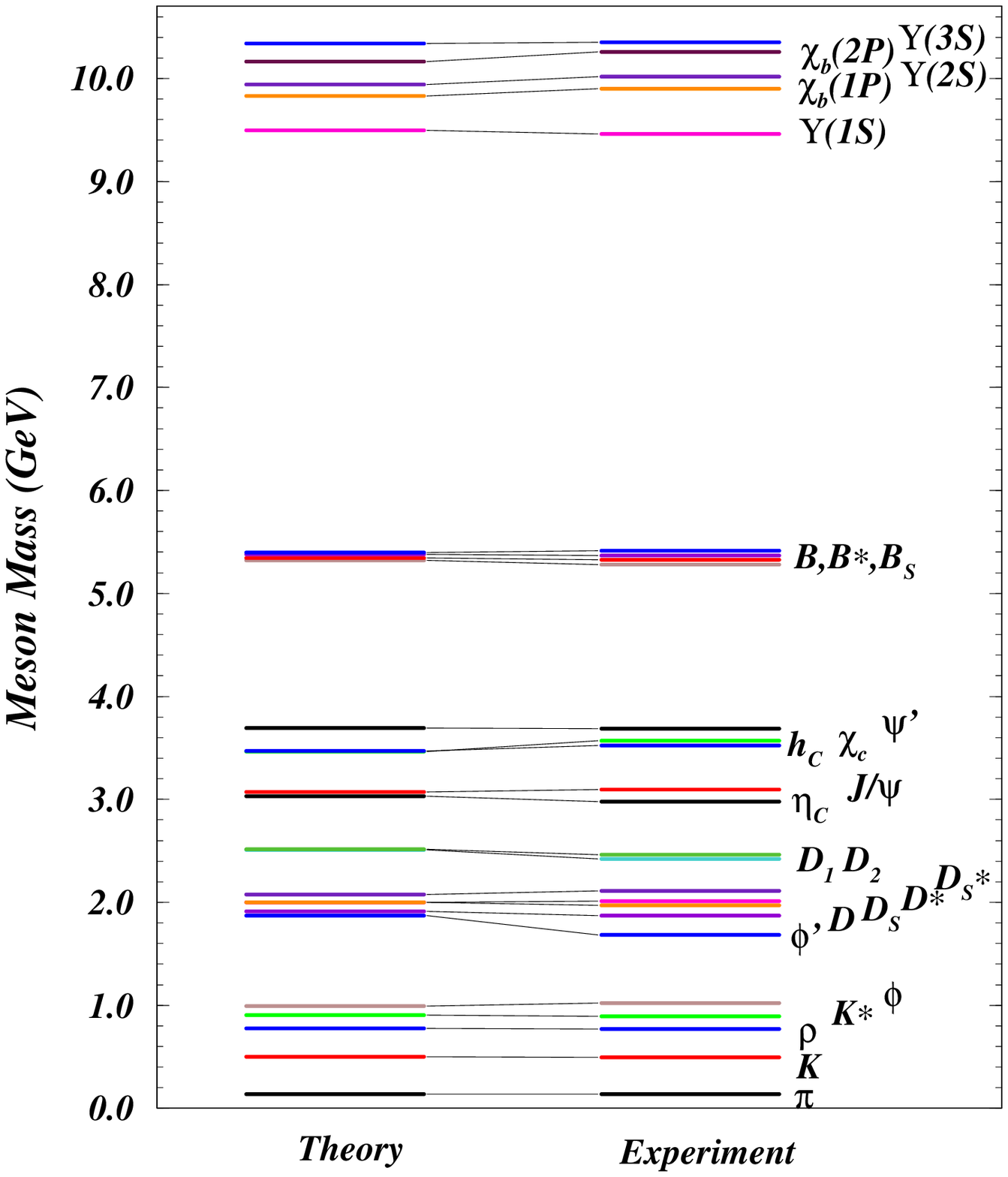}
\vspace*{8.8cm}\hspace*{3cm}
\begin{minipage}[t]{10cm}
\noindent {\bf Fig.\ 3}.  {Comparison of experimental energy levels
and theoretical energy levels calculated with the parameter set of
Eq.\ (\ref{eq:par}).}
\end{minipage}
\vskip 4truemm
\noindent 

For this study we assumed a running coupling constant combined with an
otherwise fairly conventional quark model parameter set, given by
\begin{eqnarray}
\label{eq:par}
\alpha_s(Q^2)&=&{12 \pi \over (33-2n_f) \ln(A+Q^2/B^2)},
~~~A=10,~~~B=0.31 {\rm ~GeV},
 \nonumber\\
b&=&0.18 {\rm ~GeV}^2,~~~\sigma=0.897 {\rm~GeV}, 
~~~m_u=m_d=0.334 {\rm ~ GeV}, 
\nonumber\\
m_s&=&0.575 {\rm~~GeV},~~m_c=1.776 {\rm
~GeV}, ~~m_b=5.102 {\rm~~GeV},
\nonumber \\
V_{\rm con}&=&-0.620 {\rm ~GeV}.
\end{eqnarray} 
We identified the scale $Q^2$ in the running coupling constant with
the square of the invariant mass of the interacting constituents,
$Q^2=s_{ij}$.  This set of parameters leads to a meson spectrum which
is reasonably close to experiment (see Fig.\ 3); the theoretical
masses and wavefunctions are given in Appendix A.  The parameter set
used earlier in \cite{Won00a} is similar to this set but it has a
fixed strong coupling constant.  An alternative set of quark model
parameters, without a running coupling constant, was used for
comparison.  This second set was $\alpha_s=0.594$, $b=0.162 {\rm
~GeV}^2$, $\sigma=0.897 {\rm~GeV}$, $m_u=m_d=0.335 {\rm ~ GeV}$, and
$m_c=1.6 {\rm ~GeV}$, and a flavor-dependent $V_{\rm con}$.

Having obtained the set of coefficients $\{a_n\}$ for each initial and
final meson, we can proceed to the calculation of the scattering
amplitudes $T_{fi}$ and the dissociation cross sections $\sigma_{fi}$.
Some explanation of the evaluation of our (somewhat arbitrary) choice
of momentum scale $Q^2 = s_{ij}$ in the running coupling constant
$\alpha(Q^2)$ in Eq.\ (\ref{eq:par}) for the scattering problem is
appropriate.  For a reaction process involving the interaction of
constituent $i=a$ in meson $A(aa')$ with $j=b$ in meson $B(bb')$, we
can determine the invariant mass squared $s_{ij}$ of $a$ and $b$ as
follows: Constituent $a$ carries a fraction $x_+$ of the forward
light-cone momentum of the initial meson $A$, and $b$ carries a
fraction $x_-$ of the backward light-cone momentum of initial meson
$B$.  For simplicity, we assume that the momentum fraction carried by
a constituent is proportional to its rest mass, which is exact in the
weak binding limit;
\begin{eqnarray}
x_+ = {m_a \over m_a + m_{a'}},
\end{eqnarray}
\begin{eqnarray}
x_- = {m_b \over m_b + m_{b'}}.
\end{eqnarray}
Assuming also that constituents $a$ and $b$ are on mass shell,
their momenta are \cite{Won94}:
\begin{eqnarray}
{ a_0 \choose a_z} = {1\over 2} \left [ x_+  ( A_0 + A_z) 
\pm { m_a^2  \over  ( A_0 + A_z) } \right ]
\end{eqnarray}
and
\begin{eqnarray}
{ b_0 \choose -b_z} = {1\over 2} \left [ x_-  ( B_0 - B_z) 
\pm { m_b^2  \over  ( B_0 - B_z) }\right ] ,
\end{eqnarray}
and the invariant mass of $a$ and $b$ is then given by 
\begin{eqnarray}
s_{ab}=(a+b)^2=(a_0+b_0)^2-(a_z+b_z)^2.
\end{eqnarray}
We identify this with the argument $Q^2$ of 
the running coupling constant
$\alpha_s(Q^2)$ in Eq.\ (\ref{eq:par}).

In Fig.\ 4 we show a test of the accuracy of this scattering model
with experimental data in an analogous low-energy reaction, $I=2$
$\pi\pi$ scattering. The prediction for the dominant $S$-wave phase
shift is shown together with the data of Hoogland {\it et al.} \
\cite{Hoo77}. Note that this is {\it not} a fit; all the parameters
are determined by meson spectroscopy, which fixes the interquark
forces and wavefunctions that are then used to calculate the
meson-meson scattering amplitude.

\vspace*{2.5cm}
\epsfxsize=300pt
\includegraphics{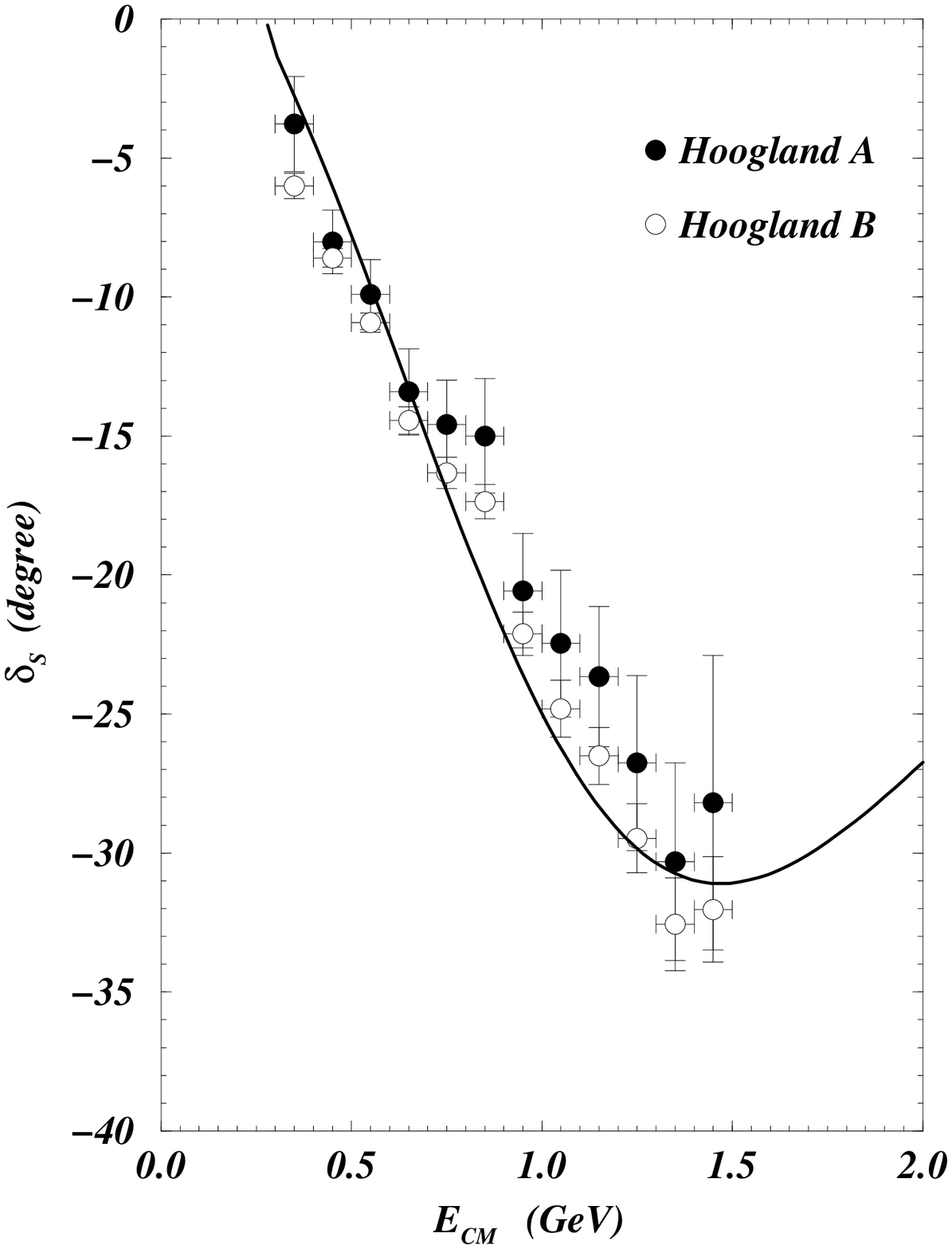}
\vspace*{10.1cm}\hspace*{3cm}
\begin{minipage}[t]{10cm}
\noindent {\bf Fig.\ 4}.  {Comparison of our theoretical $I=2$ $\pi\pi$ 
phase shift (solid curve) with the data of 
Hoogland {\it et al.} \ \cite{Hoo77}.}
\end{minipage}
\vskip 4truemm
\noindent 

\section{A Test of `Post-Prior' Equivalence in $\pi$+$J/\psi$ Dissociation}

Fig.\ 1 shows the `prior' meson-meson scattering diagrams, in which
the interactions occur before constituent interchange.  There is a
corresponding `post' set of diagrams, in which the interactions occur
after constituent interchange, as shown in Fig.\ 2.  As we noted
earlier, these are equivalent descriptions of the transition matrix
element, and should give identical cross sections.

In nonrelativistic scattering theory the post and prior results can be
formally proven to be equivalent, so that the two theoretical cross
sections are indeed identical. This proof requires that the
interaction $V$ used to determine the external meson wavefunctions be
identical to the interaction used in the evaluation of the scattering
amplitude.  Numerical confirmation of this post-prior equivalence
constitutes a very nontrivial test of the accuracy of the numerical
procedures used both in determining the bound state wavefunction and
in evaluating the complete meson-meson scattering amplitude.  This
post-prior equivalence was discussed in detail and demonstrated
numerically by Swanson for $\pi \pi \to \rho \rho $ scattering in
Ref.\cite{Swa92}, where its relevance to establishing Hermitian
effective scattering potentials was shown.

To test post-prior equivalence in our $J/\psi$ dissociation
calculations (in the nonrelativistic formalism), we have carried out
the evaluation of the cross section using both post and prior
formalisms.  Of necessity we assumed nonrelativistic kinematics and
theoretical masses to determine the external meson momenta
$|{\bbox{A}}|$ and $|{\bbox{C}}|$, which appear in the overlap
integrals.  The post-prior equivalence holds if the coupling constant
does not depend on energy.  We are well advised to use a set of
parameters with a fixed running coupling constant $\alpha_s$ for this
test.  We used therefore the parameter set \cite{Won00a}
$\alpha_s=0.58$, $b=0.18$ GeV$^2$, $\sigma$=0.897 GeV, $m_{u,d}=0.345$
GeV, $m_c=1.931$ GeV, and $V_{\rm con}=-0.612$ GeV, which are close to
standard values and were chosen because they reproduce the physical
masses of the $\pi$, $D$, $D^*$, $J/\psi$ and $\psi'$ rather well.
Fig.\ 5$a$ shows the dissociation cross sections for $\pi$+$\psi$
collisions as a function of the initial kinetic energy $E_{KE}$
measured in the center-of-mass frame, defined as $E_{KE}=\sqrt{s}-m_A
-m_B$ where $m_A$ and $m_B$ are the masses of the incident mesons.
The differences between the post and prior results in Fig.\ 5$a$ are
indeed rather small, which confirms post-prior equivalence assuming
non-relativistic dynamics.  (The residual discrepancy is mainly due to
our use of a truncated basis in expanding the meson wavefunctions.)
Fig.\ 5$b$ shows the corresponding results for $\pi$+$\psi'$
dissociation.  In this case there is greater sensitivity to the
approximate $\psi'$ wavefunction, due to large cancellations in the
numerical integration of a radially-excited wavefunction.

\vspace*{-0.5cm}
\epsfxsize=300pt
\includegraphics{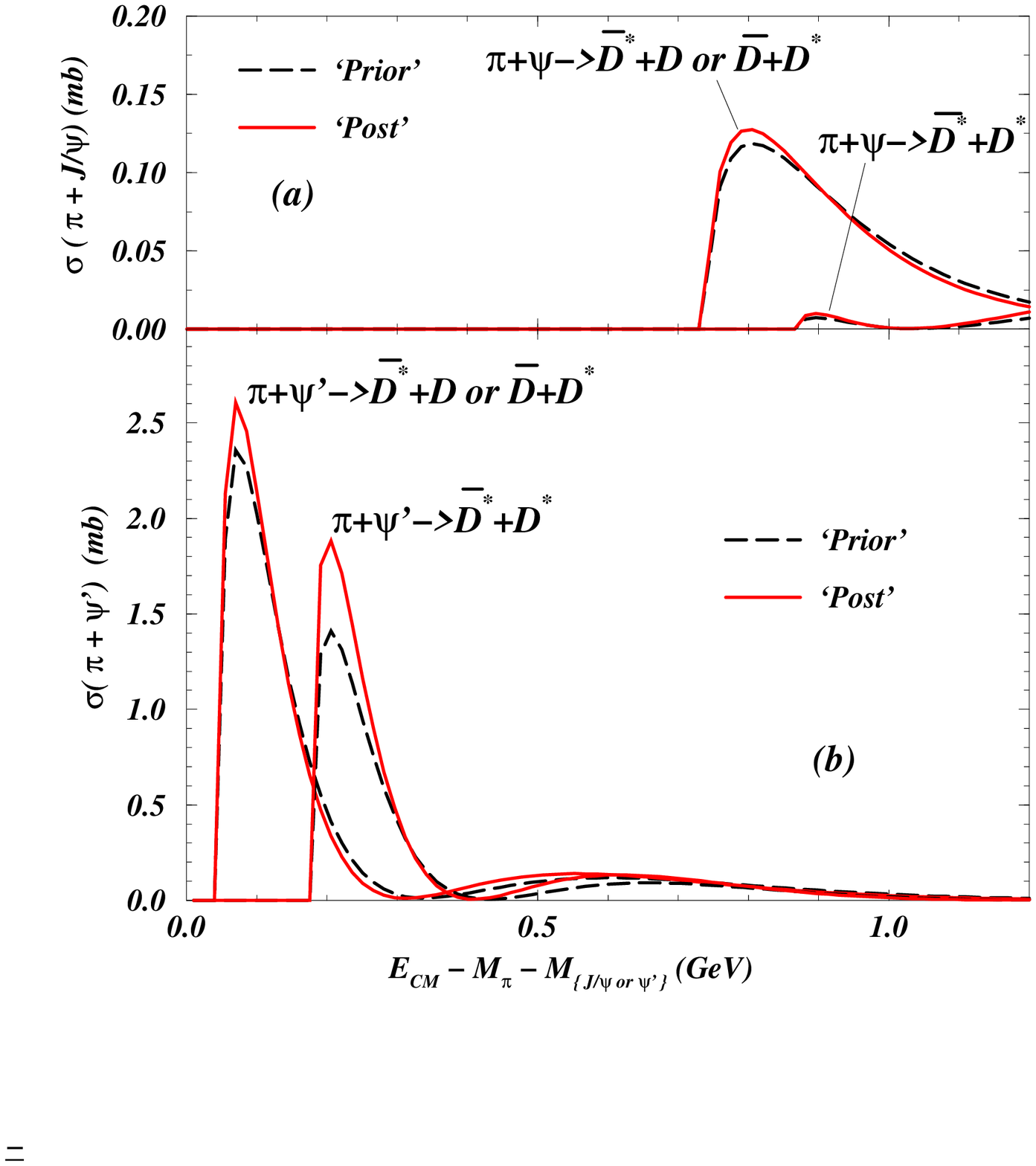}
\vspace*{10.1cm}\hspace*{1.5cm}
\begin{minipage}[t]{10cm}
\noindent {{\bf Fig.\ 5}.  Comparison of $\pi+J/\psi$ and $\pi+ \psi'$
cross sections derived using post and prior formalisms.}
\end{minipage}
\vspace*{0.3cm}
\noindent 

In the comparisons shown in Fig.\ 5 we have used theoretical masses
for the mesons; the proof of post-prior equivalence makes use of the
theoretical bound-state masses from the Schr\"odinger equation with
the given potential, rather than the experimental ones. Our
theoretical masses are close to experiment but are not exact, as is
evident in Fig.\ 3.  If one instead assumes experimental values for
the meson masses, there will be additional post-prior discrepancies in
our cross section calculations.  In our subsequent cross section
calculations we do assume experimental masses in order to reproduce
correct thresholds; the consequence is a systematic uncertainty in the
cross sections which may be estimated by comparing post and prior
predictions.

It should be noted that `post-prior equivalence' had only been proven
for bound-state scattering in non-relativistic quantum mechanics
\cite{Sch68}.  Recently an extension of this proof to scattering in a
relativistic generalization of quantum mechanics was given by Wong and
Crater \cite{Won00c}, using Dirac's constraint dynamics.  A full
relativistic treatment of this problem will involve the derivation of
relativistic two-body wavefunctions and Wigner spin rotations, which
is beyond the scope of the present investigation.  We will adopt an
intermediate approach and assume relativistic kinematics for the
initial and final mesons and use relativistic phase space; in
consequence we find different `post' and `prior' cross sections in
general.  Here we will take the mean value of the `post' and `prior'
results as our estimated cross section, separate `post' and `prior'
cross sections will be shown as an indication of our systematic
uncertainty.

\section{Cross sections for $J/\psi$ and $\psi'$ dissociation}

Depending on the incident energies, dissociation of the $J/\psi$ and
$\psi'$ by pions can lead to many different exclusive final states.
There are several selection rules that eliminate or suppress many of
the {\it a priori} possible final channels, given our simple
quark-model Hamiltonian and Born-order scattering amplitudes.
Considerable simplification follows from the fact that our model
Hamiltonian conserves total spin $S_{tot}$.  Since the $J/\psi$ and
$\psi'$ have $S=1$ and pions have $S=0$, the initial and final states
in $\pi$+$J/\psi$ and $\pi$+$\psi'$ collisions must both have
$S_{tot}=1$; this forbids $D\bar D$ final states. With increasing
invariant mass we next encounter the final states $D \bar D^*$, $ D^*
\bar D$ and ${D^*} \bar D^*$.  In Fig.\ 6 we show the exclusive
dissociation cross sections for $\pi$+$J/\psi$ and $\pi$+$\psi'$
collisions into these first few allowed final states.  The total
dissociation cross section, which is the sum of the exclusive cross
sections, is shown as a solid line.  Our estimate is the mean of the
total `post' and `prior' cross sections, which are also shown in Fig.\
6.  The estimated range of uncertainty, due to the post-prior
discrepancy and to parameter variations, is shown as a shaded band.

The $J/\psi$ dissociation processes $\pi$+$J/\psi\to D^*\bar D$ or
$D\bar D^*$ have a threshold of about $E_{KE}$=0.65 GeV and the total
dissociation cross section reaches approximately 1~mb not far above
threshold (Fig.\ 6$a$).  This $\pi+J/\psi$ cross section is
considerably smaller than the peak value of about 7~mb found by
Martins $et~al.$, largely due to their assumption of a
color-independent confining interaction.

The threshold for $\pi+\psi'\rightarrow{D}\bar D^*+{ D^*} \bar D $ is
only about $E_{KE} = 0.05$ GeV, and in consequence the $\psi'$ cross
sections are much larger near threshold.  The total $\pi$+$\psi'$
dissociation cross section has maxima of $\approx 4.2(0.5)$~mb and
$\approx 2.8(0.5)$~mb at $E_{KE}\approx 0.1$ GeV and $\approx 0.22$
GeV respectively (Fig.\ 6$b$).  It is interesting that the exclusive
$\psi'$ dissociation cross sections are very small near $E_{KE} = 0.3$
(for the $D{\bar D}^*$ final state) and 0.4~GeV (for the $D^*{\bar
D}^*$ final state), due to almost complete destructive interference
between the diagrams.

\vspace*{10.1cm}
\hskip 1cm
\epsfxsize=300pt
\includegraphics{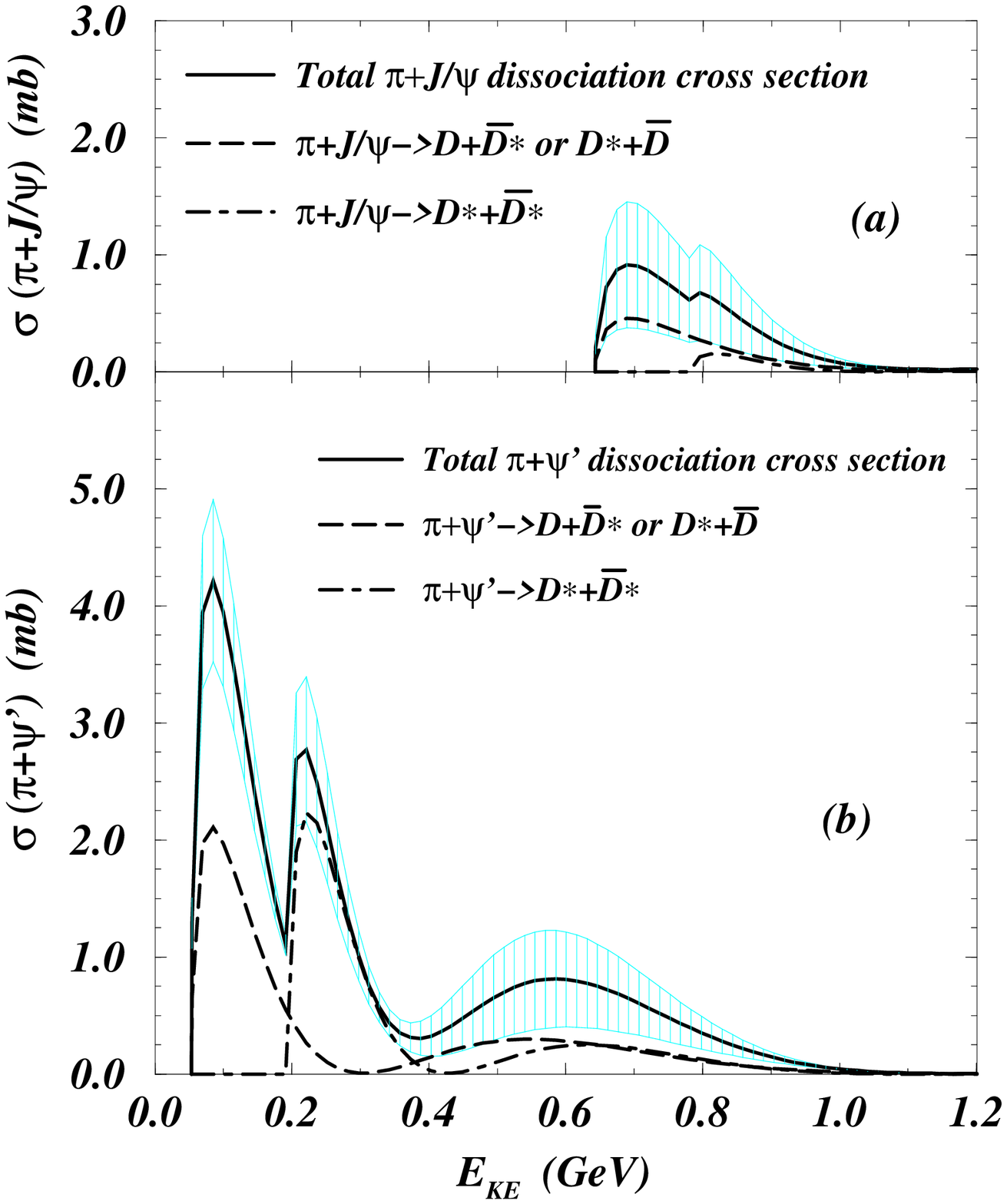}
\vspace*{0.8cm}\hspace*{1.5cm}
\begin{minipage}[t]{12cm}
\noindent {{\bf Fig.\ 6}.  Exclusive and total dissociation cross
sections for $\pi$+$J/\psi$ (Fig.\ 6$a$) and $\pi$+$\psi'$ (Fig.\
6$b$).  In each panel the solid curve gives our estimated total cross
section, which is the mean of the `prior' and `post' results.}
\end{minipage}
\vskip 2truemm
\noindent 

The relative importance of the various terms in the Hamiltonian in
these dissociation amplitudes is of course a very interesting
question.  Unfortunately it is also somewhat ambiguous, because the
individual terms differ between post and prior formalisms; only the
sum of all terms is formalism independent in non-relativistic quantum
mechanics.  We find that the spin-spin interaction makes the dominant
contribution to $\pi$+$J/\psi$ dissociation in the prior formalism;
$\pi$+$\psi'$ dissociation in contrast is dominated by the linear
confining interaction.  In the post formalism we find that both
$\pi$+$J/\psi$ and $\pi$+$\psi'$ are dominated by the spin-spin
interaction. In all these cases the color-Coulomb contribution is
subdominant.

Our results have interesting consequences for the survival of $J/\psi$
and $\psi'$ mesons propagating in a gas of pions.  The pions produced
in a heavy ion collision have a roughly thermal distribution, with
$T\approx 200$~MeV in the nucleus-nucleus center-of-mass system,
whereas heavy quarkonia such as the $J/\psi$ and $\psi'$ are produced
approximately at rest.  Thus the relative kinetic energy is typically
a few hundred MeV.  This is below the $\pi$+$J/\psi$ dissociation
threshold, but above that of $\pi$+$\psi'$, and in consequence we
expect $\pi$+$\psi'$ collisions to deplete the $\psi'$ population much
more effectively than $\pi$+$J/\psi$ collisions reduce the initial
$J/\psi$ population. The weakness of $\pi$+$J/\psi$ dissociation is
further reduced by the small cross section we find for the
$\pi$+$J/\psi$ relative to $\pi$+$\psi'$.

Next we consider the very interesting exothermic collisions of
charmonia with light vector mesons, specifically $\rho$+$J/\psi$ and
$\rho$+$\psi'$.  Since the $\rho$ meson has $S=1$, the total spin of
the $\rho + \{ J/\psi {\rm ~or~} \psi'\}$ system can be
$S_{tot}=\{0,1,$ and $2\}$.  This $S_{tot}$ is conserved in our model,
and so must agree with the $S_{tot}$ of the final state. The low-lying
open charm final states are $ D \bar D$ with $S_{tot}=0$, $ D {\bar
D^*}$ and $ {D^*} \bar D$ with $S_{tot}=1$, and $D^* {\bar D^*}$ with
$S_{tot}=\{ 0, 1,$ and $2\}$.  The contribution of transitions to
radially- and orbitally-excited final states will be considered in
future work \cite{Bar01}.

\vspace*{3.1cm}
\epsfxsize=300pt
\includegraphics{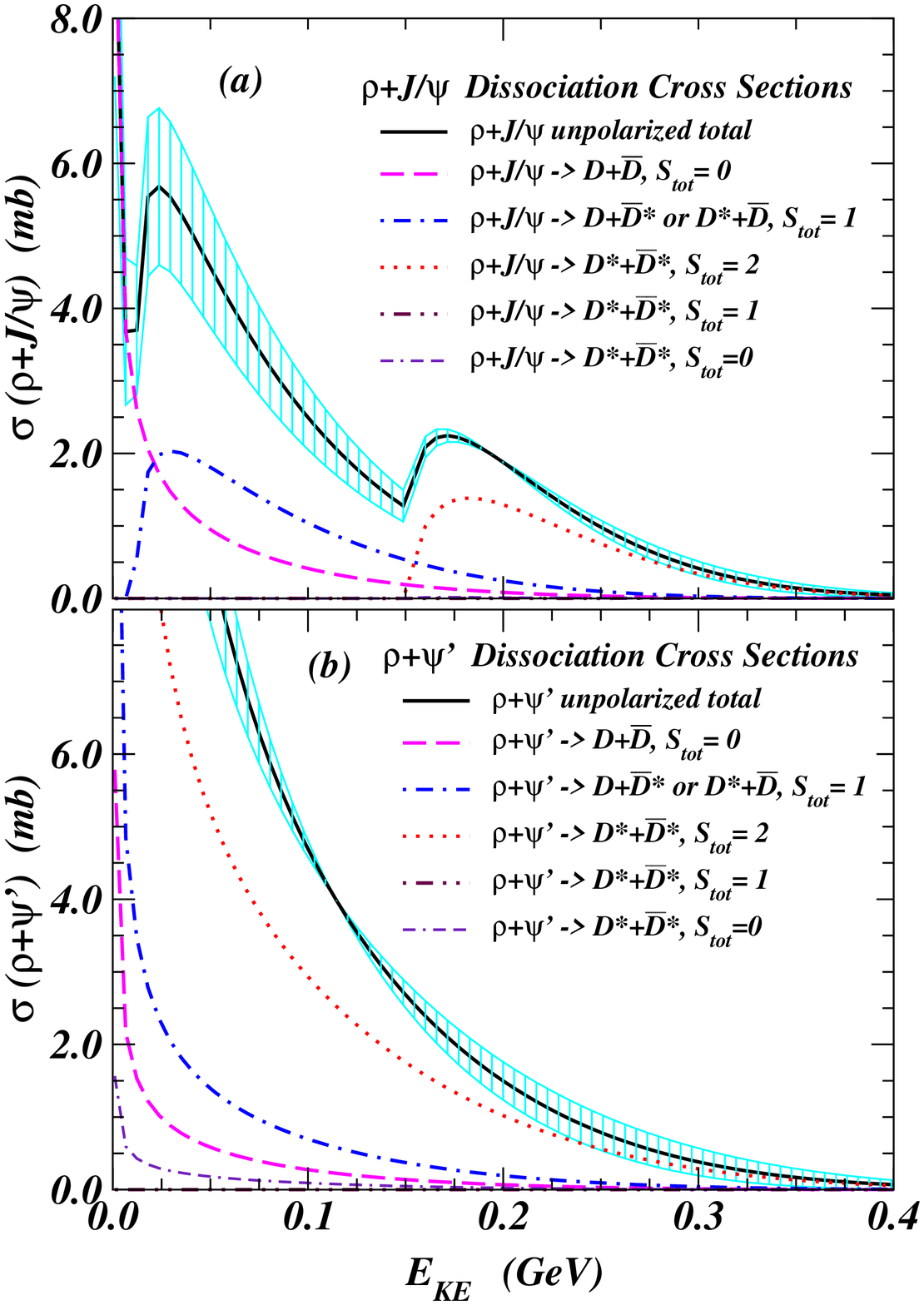}
\vspace*{9.6cm}\hspace*{1.5cm}
\begin{minipage}[t]{12cm}
\vskip 1cm
\noindent {\bf Fig.\ 7}.  { The total unpolarized dissociation cross
sections are shown as the solid curves for $\rho$+$J/\psi$ in Fig.\
7$a$ and for $\rho$+$\psi'$ in Fig.\ 7$b$.  Unpolarized exclusive
dissociation cross sections $\sigma^{\rm unpol}_f(S_{tot})$ for different
final states and different $S_{tot}$ are also shown.  }
\end{minipage}
\vskip 4truemm
\noindent 
\vspace*{0.2cm}

In the collision of an unpolarized $\rho$ with an unpolarized
$J/\psi$, the total dissociation cross section comprises of
contributions $\sigma_f^{\rm unpol}(S_{tot})$ from different final states
$f$ and different total spin values $S_{tot}$ of the system,
\begin{eqnarray}
\label{eq:unp}
\sigma_{tot}^{\rm unpol}=
\sum_f \sum_{S_{tot}} \sigma_f^{\rm unpol} (S_{tot}),
\end{eqnarray} 
where for this case with $L_A=0$ and $S_B=S_\rho\ne 0$
we can deduce from Eq.\ (\ref{eq:unpola}) the following
relationship 
\begin{eqnarray}
\label{eq:unp1}
\sigma_f^{\rm unpol} (S_{tot})=
{(2S_{tot}+1) \over (2S_\rho
+1)(2S_{J/\psi}+1)} 
\sigma_f (S_{tot}),
\end{eqnarray} 
and $\sigma_f(S_{tot})$ is the dissociation cross section for the
final state $f$ when the initial two-meson system is prepared with a
total spin $S_{tot}$.  [In our earlier work \cite{Won00a} for
$\rho+J/\psi$ and $\rho+\psi'$ collisions, $\sigma_f(S_{tot})$ results
were presented and the total cross section of $\sigma_{tot}=\sum_f
\sum_{S_{tot}} \sigma(S_{tot})$ was evaluated.  However, for the
collision of unpolarized mesons, one should use the unpolarized total
dissociation cross section as given by Eqs.\ (\ref{eq:unp}) and
(\ref{eq:unp1}).].

The unpolarized total $\rho + J/\psi$ dissociation cross section are
shown in Fig.\ 7$a$.  The exclusive cross sections $\sigma_f^{\rm
unpol} (S_{tot})$ for dissociating into various final states in an
unpolarized collision are also exhibited.  The reaction $\rho + J/\psi
\to D \bar D $ is exothermic, so the cross section $\sigma_{D \bar
D}^{\rm unpol}(S_{tot}=0)$ diverges as $1/\sqrt{E_{KE}}$ as we
approach threshold.  For other $\rho + J/\psi$ exclusive final states
the thresholds lie at sufficiently higher energies to be endothermic.
We find an unpolarized total $\rho + J/\psi$ dissociation cross
section of 2.4(0.4)~mb at $E_{KE}=0.1$~GeV, which has decreased to
about 1.9~mb at $E_{KE}=0.2$~GeV.  At very low kinetic energies, the
contribution to the dissociation of $J/\psi$ by $\rho$ comes from the
$D\bar D (S_{tot}=0)$ final state.  At slightly higher energy, it
comes mainly from $D\bar D^*$ and $D^* \bar D(S_{tot}=1)$ final
states. At $E_{KE}$ about 0.2 GeV, it comes dominantly from the $D^*
\bar D^* (S_{tot}=2)$ final state.

We have carried out similar calculations for $\rho+\psi'$ collisions,
and the results are shown in Fig.\ 7$b$.  In this case the reactions
$\rho + \psi' \to D \bar D, D {\bar D^*}, D^* {\bar D}$ and $ D^* \bar
D^*$ are all exothermic, so all these exclusive cross sections
$\sigma_f^{\rm unpol} (S_{tot})$ diverge as $1/\sqrt{E_{KE}}$ as we
approach threshold.  The dominant contribution to the dissociation
comes from the $D^*\bar D^*(S_{tot}=2)$ final state.  The total
unpolarized $\rho + \psi'$ dissociation cross section falls from
4.5(0.1)~mb at $E_{KE} = 0.1$~GeV to 1.5(0.3)~mb at $E_{KE} = 0.2
$~GeV and 0.4(0.2)~mb at $E_{KE} = 0.3$~GeV.

In $\rho$+$J/\psi$ dissociation the dominant scattering contribution
in the prior formalism is due to the linear interaction.  In the post
formalism the dominant contribution arises from the color-Coulomb and
linear interactions at energies $E_{KE} < 0.1$ GeV, from color-Coulomb
at 0.1 GeV$< E_{KE} < $0.4 GeV, and from the spin-spin interaction at
$E_{KE} > 0.4$ GeV.

We turn next to the dissociation of $J/\psi$ and $\psi'$ in collision
with $K$.  Our predictions for $K+J/\psi$ and $K+\psi'$ dissociation
cross sections are shown in Fig.\ 8.  The $K+J/\psi$ process has a
threshold kinetic energy of about 0.4~GeV, and the maximum cross
section is about 0.7~mb.  In $K$+$\psi'$ dissociation the reactions
are exothermic for the allowed final states $D_s+\bar D^*, D_s^*+\bar
D$ and $ D_s^*+\bar D^*$.  The total $K$+$\psi'$ dissociation cross
section, shown in Fig.\ 8$b$, is about 1~mb at $E_{KE}\sim 0.4$~GeV,
and also diverges as $1/\sqrt{E_{KE}}$ as we approach threshold.

\vspace*{2.0cm}
\epsfxsize=300pt
\includegraphics{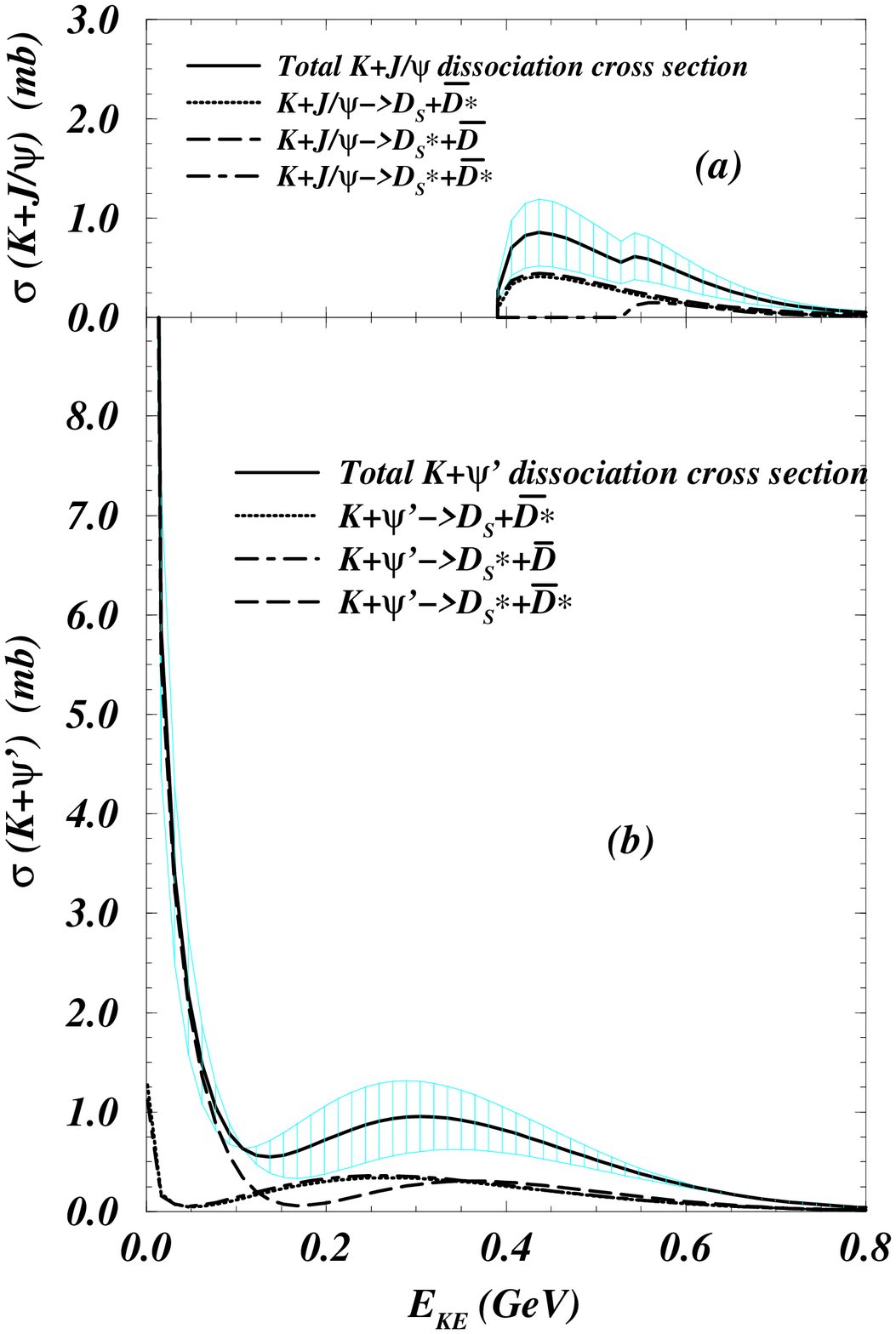}
\vspace*{10.1cm}\hspace*{1.5cm}
\begin{minipage}[t]{12cm}
\noindent {\bf Fig.\ 8}.  {Dissociation cross sections for
$K$+$J/\psi$ (Fig.\ 8$a$) and $K$+$\psi'$ (Fig.\ 8$b$).}
\end{minipage}
\vskip 4truemm
\noindent 
\vspace*{-0.4cm}

The model of Barnes and Swanson has thus far received experimental
support from extensive comparisons with light hadron scattering data
($I=2$ $\pi$$\pi$ \cite{Bar92}, $I=3/2$ $K\pi$ \cite {Kpi}, $I=0,1$
$KN$ \cite{KN}).  The new results of $J/\psi$ and $\psi'$ dissociation
in collision with $\pi$, $\rho$, and $K$ need to confront experimental
data through an examination of their implications on $J/\psi$ and
$\psi'$ suppression in high-energy heavy-ion collisions.  The extent
to which the observed anomalous suppression in Pb+Pb collisions is due
to the dissociation $J/\psi$ by $\pi$, $\rho$, and $K$ will require
further quantitative study.  $J/\psi$ suppression by these mesons must
be incorporated and subtracted in order to identify the suppression of
$J/\psi$ production by the quark-gluon plasma.  It is therefore
important to carry out a detailed simulation of $J/\psi$ absorption
using cross sections obtained here in order to understand the nature
of $J/\psi$ suppression in Pb+Pb collisions and to provide a test of
the theoretical $J/\psi$ dissociation cross sections.

\section{Cross sections for $\Upsilon$ and $\Upsilon'$ dissociation.}

It has been noted that the suppression of the production rates of
$b\bar b$ mesons, such as the $\Upsilon$ and $\Upsilon'$, may also be
useful as a signal for QGP production (see Ref.\ \cite{Lin01} and
references cited therein).  It is of interest to calculate the
$\Upsilon$ and $\Upsilon'$ dissociation cross section in collisions
with $\pi$, $\rho$, and $K$ so as to infer the effects of $\Upsilon$
and $\Upsilon'$ suppression by hadronic produced particles.

\vspace*{1.0cm}
\epsfxsize=300pt
\includegraphics{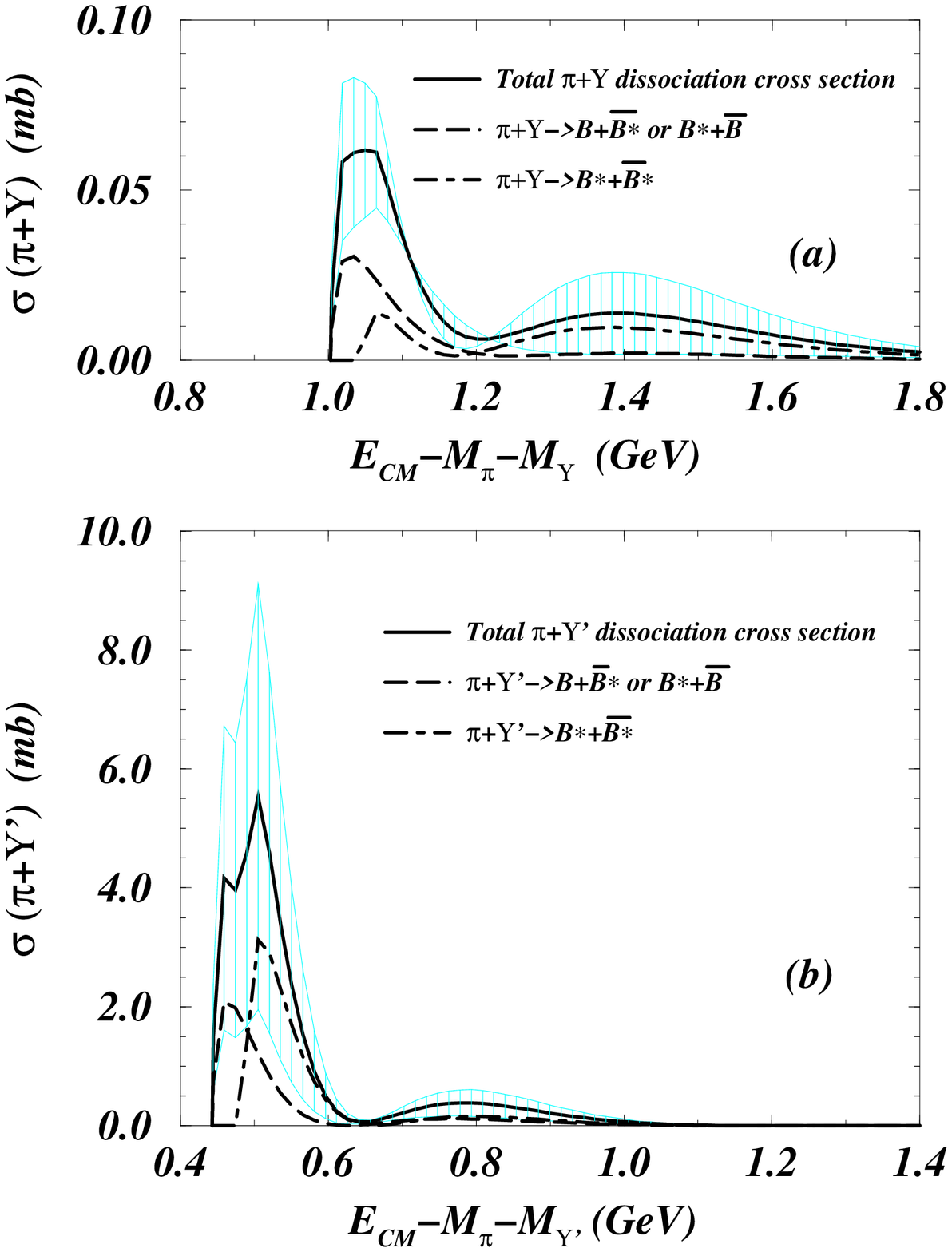}
\vspace*{10.1cm}\hspace*{1.5cm}
\begin{minipage}[t]{12cm}
\noindent {\bf Fig.\ 9}.  {The dissociation cross sections for
$\Upsilon$ (Fig.\ 9$a$) and $\Upsilon'$ (Fig.\ 9$b$) in collision
with $\pi$. Note that the scales of Figs.\ 9$a$
and 9$b$ are different.}
\end{minipage}
\vskip 4truemm
\noindent 
\vspace*{0.2cm}

We show in Fig.\ 9$a$ the total cross section for the dissociation of
the $\Upsilon$ in collision with $\pi$.  The threshold is at
$E_{KE}\sim 1$ GeV, and the maximum cross section is about 0.05 mb.
The small cross section arises from the combined effects of a large
threshold and a small value of the strong interaction coupling
constant.  We show the dissociation cross section for $\Upsilon'$ in
collision with $\pi$ in Fig.\ 9$b$.  It has a threshold of $E_{KE}\sim
0.45$ GeV, and the peak cross section is about 5 mb at $E_{KE}\sim
0.55$ GeV.  This dissociation cross section is much larger than that
for the dissociation of $\Upsilon$ by $\pi$.

In Fig.\ 10$a$ we show the unpolarized cross section for the
dissociation of $\Upsilon$ in collision with $\rho$.  Unpolarized
exclusive dissociation cross sections $\sigma^{\rm unpol}_f(S_{tot})$ for
different final states and different $S_{tot}$ are also shown.  The
reaction process is endothermic with a threshold at $E_{KE}\sim 0.3$
GeV and a peak cross section of 0.15 mb at $E_{KE}\sim 0.45$ GeV.  In
Fig.\ 10$b$ we show the dissociation cross section for $\Upsilon'$ in
collision with $\rho$.  The reactions are exothermic and the total
dissociation cross section behaves as $1/\sqrt{E_{KE}}$ near
$E_{KE}\sim 0$ and decreases to about 1.6 mb at $E_{KE}\sim 0.2$ GeV.

\vspace*{3.1cm}
\epsfxsize=300pt
\includegraphics{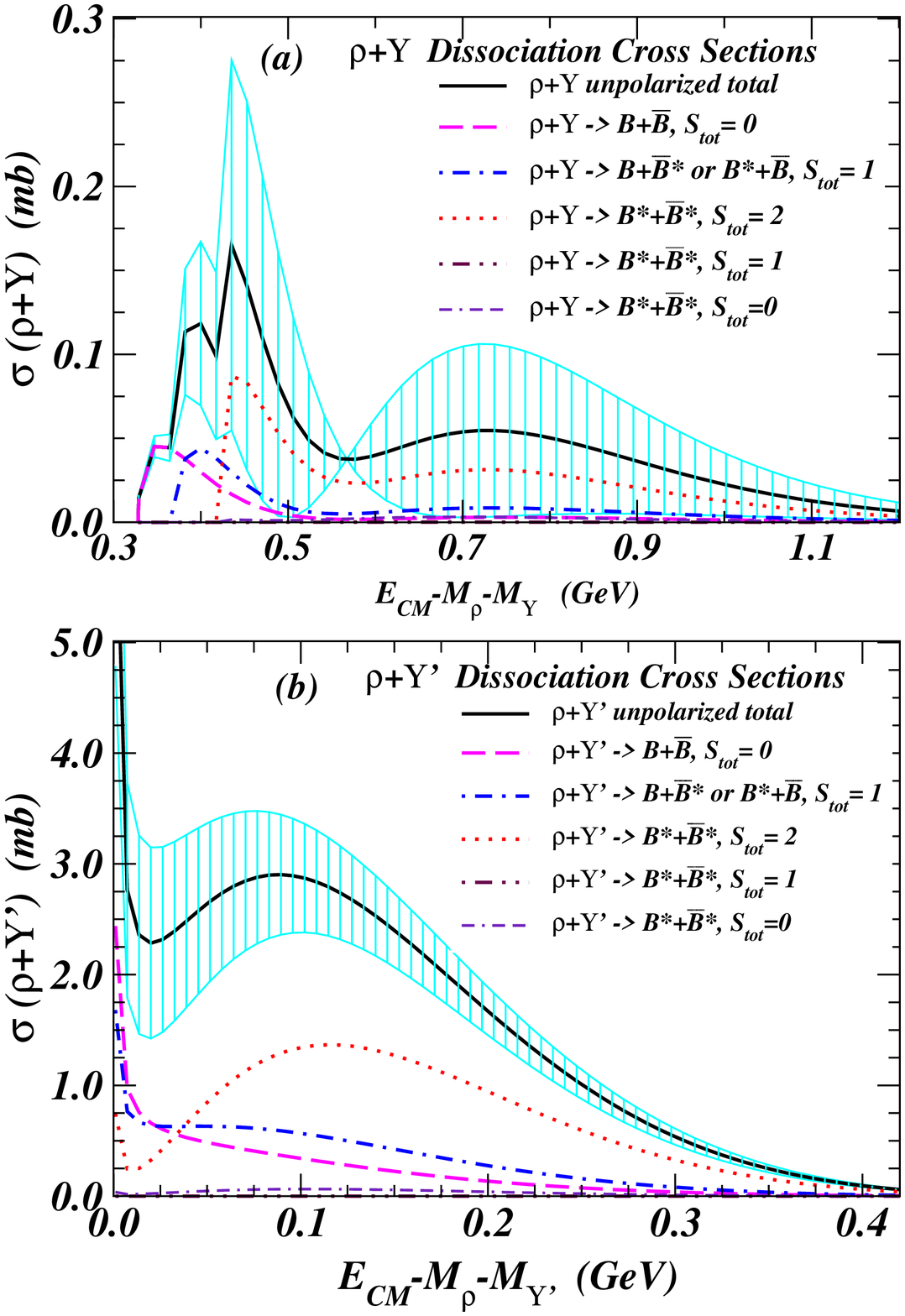}
\vspace*{10.6cm}\hspace*{1.5cm}
\begin{minipage}[t]{12cm}
\noindent {\bf Fig.\ 10}.  {Unpolarized total dissociation cross
sections and unpolarized exclusive dissociation cross sections
$\sigma_f^{\rm unpol}(S_{tot})$ for $\rho+\Upsilon$ (Fig.\ 10$a$) and
$\rho+\Upsilon'$ (Fig.\ 10$b$) for various channels and total spin
$S_{tot}$.  Note that Figs.\ 10$a$ and 10$b$ have different scales.}
\end{minipage}
\vskip 4truemm
\noindent 
\vspace*{0.2cm}

In Fig.\ 11$a$ we show the cross section for the dissociation of
$\Upsilon$ by $K$.  The threshold is at $E_{KE}\sim 0.75$ GeV, with a
peak total cross section of about 0.05 mb at $E_{KE}\sim 0.85$ GeV.
We show in Fig.\ 11$b$ the dissociation cross section of $\Upsilon'$
by $K$.  The threshold is at $E_{KE}\sim 0.2$ GeV, with a peak total
cross section of about 2 mb at $E_{KE}\sim 0.25$ GeV.

\vspace*{1.7cm}
\epsfxsize=300pt
\includegraphics{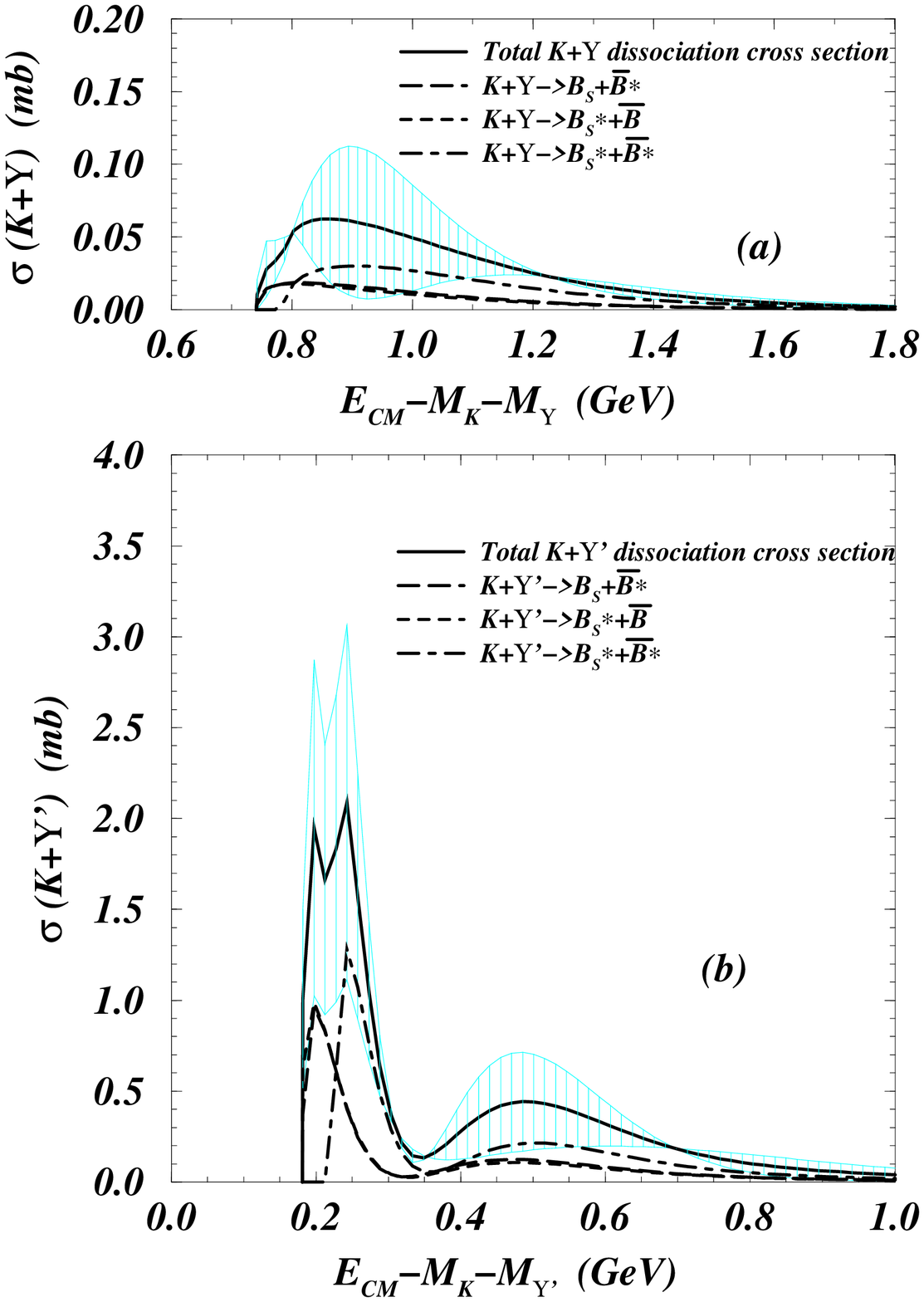}
\vspace*{10.1cm}\hspace*{1.5cm}
\begin{minipage}[t]{12cm}
\noindent {\bf Fig.\ 11}.  {The dissociation cross sections for
$K+\Upsilon'$ (Fig.\ 11$a$) and $K+\Upsilon'$ (Fig.\ 11$b$) for
various final states. Note that Figs.\ 11$a$ and 11$b$ have different
scales.}
\end{minipage}
%\vskip 4truemm
%\noindent 
%\vspace*{0.2cm}

\section{ Cross Sections for $\chi_J$ Dissociation }

We can calculate the dissociation cross sections of $\chi_J$ mesons in
collision with $\pi$, $\rho$, and $K$ using the quark-interchange
model. A $\chi_J$ meson has a spin quantum number $S=1$ and a $\pi$
has $S=0$.  The collision of a $\chi_J$ with a $\pi$ gives rise to a
system with a total spin $S_{tot}=1$.  On the other hand, the
interaction of Eq.\ (2), which leads to the dissociation, conserves
the total spin.  Therefore the lowest-energy final states are $D{\bar
D}^*$, $D^* {\bar D}$, and $D^* {\bar D}^*$.

We show the results for the dissociation cross sections of unpolarized
$\chi_{J}$ in collision with $\pi$ in Fig.\ 12.  The unpolarized
dissociation cross sections for the final states of $D {\bar D}^*$ and
$D^* {\bar D}^*$ are shown as the dotted and the dashed-dot curves
respectively.  The unpolarized total cross sections for scattering
into these lowest channels are shown as the solid curve obtained as
the mean value of the total cross section from the `prior' and `post'
formalisms.  The estimated systematic uncertainty in the total cross
section due to the post-prior discrepancy is again indicated as a band
in the figure.  The dissociation of $\chi_{0}$ by $\pi$ has a
threshold of $E_{KE} \sim 0.32$ GeV and a peak cross section of 1.53
mb at $E_{KE} \sim 0.5$ GeV (Fig.\ 12$a$).  The dissociation of
$\chi_{1}$ by $\pi$ has a threshold of $E_{KE} \sim 0.23$ GeV and a
peak dissociation cross section of 2.41 mb at $E_{KE} \sim 0.46$ GeV
(Fig.\ 12$b$).  The dissociation of $\chi_{2}$ by $\pi$ has a
threshold of $E_{KE} \sim 0.18$ GeV and a peak dissociation cross
section of about 2.98 mb at $E_{KE} \sim 0.41$ GeV (Fig.\ 12$c$).

\vspace*{14cm}
\hskip -1cm
\epsfxsize=300pt
\includegraphics{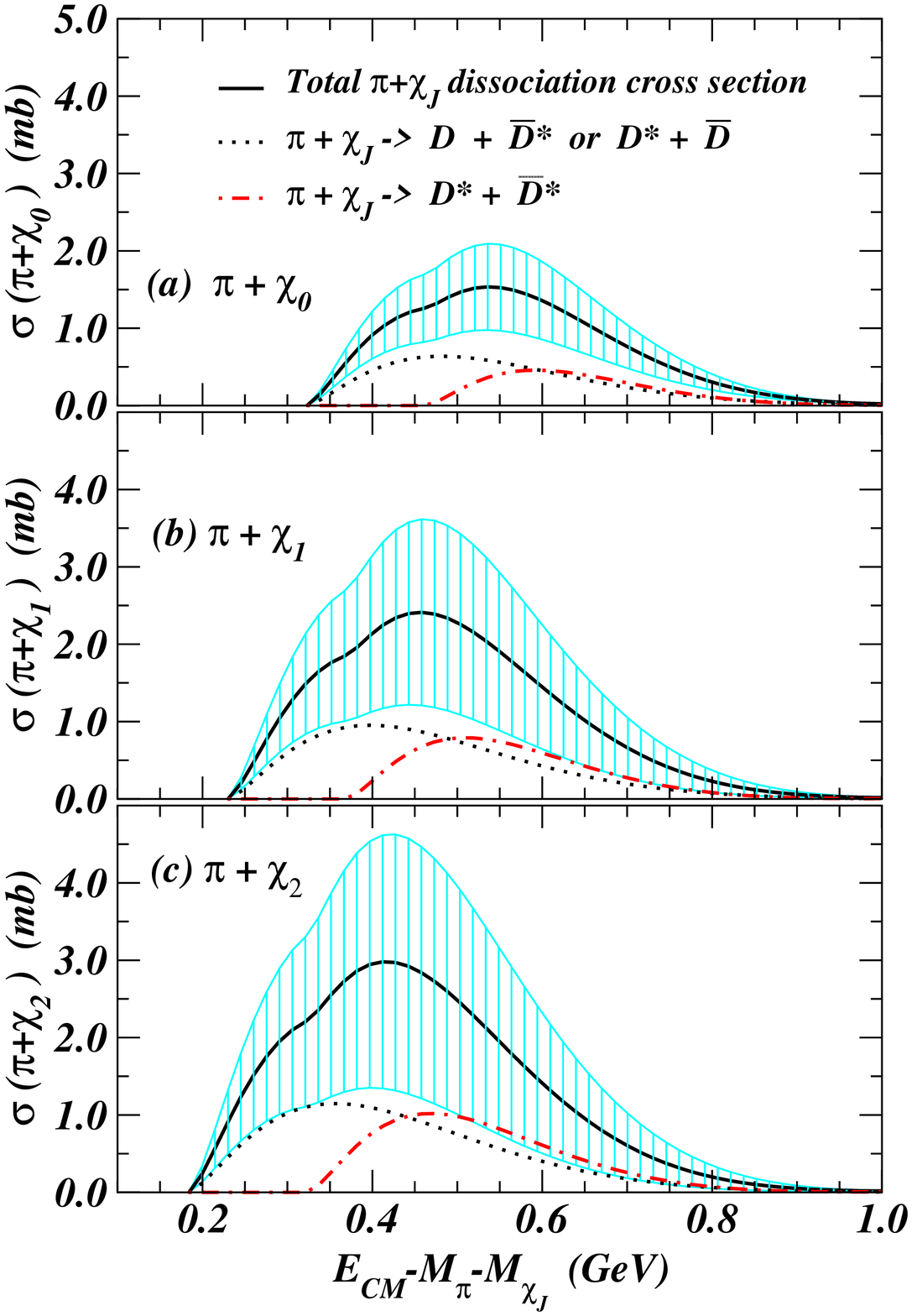}
\vspace*{0.5cm}\hspace*{1.5cm}
\begin{minipage}[t]{12cm}
\noindent {{\bf Fig.\ 12}.  Unpolarized total cross sections and
exclusive unpolarized cross sections for the dissociation of $\chi_J$
into $D + \bar D^*$ and $D^* +\bar D^*$ in collision with $\pi$.
Fig.\ 12$a$ is for $\pi+\chi_0$, Fig.\ 12$b$ for $\pi+\chi_{1}$, and
Fig.\ 12$c$ for $\pi+\chi_{2}$.}
\end{minipage}
\vskip 4truemm
\noindent 

It is interesting to note that the maximum of the unpolarized total
dissociation cross section for $\pi+\chi_{2}$ is only slightly greater
than that for $\pi+\chi_{1}$ but is nearly twice as great as the
maximum of the dissociation cross section for $\pi+\chi_0$.  This
indicates that the dissociation of $\chi_J$ is very sensitive to the
threshold value.  We found numerically that if the threshold value for
$\pi+\chi_0$ were taken to be the same as the threshold value for
$\pi+\chi_{2}$, the unpolarized dissociation cross sections would be
the same.

The dissociation amplitudes of the $\chi_{JJ_z}$ mesons in collision
with pions depend on $J_z$.  A detailed discussion of the dissociation
cross section for various $J$ and $J_z$ substates will be presented in
\cite{Bar01}.  The dependence on $J_z$ is however quite weak.  For the
same value of $J$, dissociation cross sections of $\chi_{J J_z}$ in
collision with $\pi$ vary only by a few percent for different $J_z$.

The thresholds for $\pi+\chi_J$ dissociation lie in between those of
$\pi+J/\psi$ and $\pi+\psi'$, and the maxima of the total dissociation
cross sections for the $\pi+\chi_J$ collisions are greater than that
for the $\pi+J/\psi$ collision but less than for the $\pi+\psi'$
collision.

\vspace*{14cm}
\hskip -1cm
\epsfxsize=300pt
\includegraphics{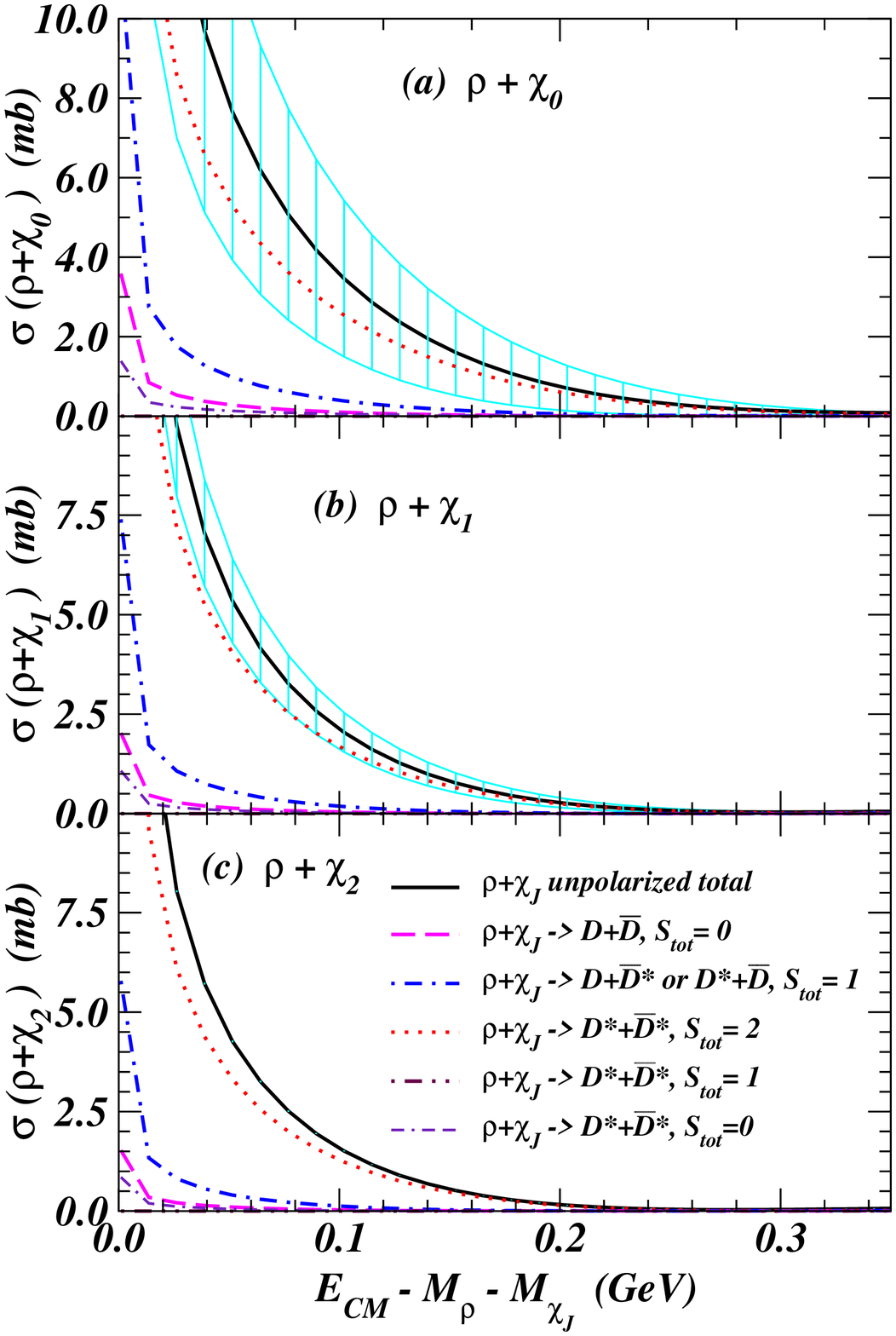}
\vspace*{0.5cm}\hspace*{1.5cm}
\begin{minipage}[t]{12cm}
\noindent {{\bf Fig.\ 13}.  Unpolarized total cross sections and
unpolarized exclusive cross sections $\sigma_f^{\rm unpol}(S_{tot})$
for the dissociation of $\chi_J$ into $D\bar D$, $D + \bar D^*$ and
$D^* +\bar D^*$ in collision with $\rho$.  Fig.\ 13$a$ is for
$\rho+\chi_0$, Fig.\ 13$b$ for $\rho+\chi_{1}$, and Fig.\ 13$c$ for
$\rho+\chi_{2}$.}
\end{minipage}
\vskip 4truemm
\noindent 

We show in Fig.\ 13 the dissociation cross sections of unpolarized
$\chi_J$ in collision with unpolarized $\rho$.  The lowest-energy
final states are $D\bar D$, $D\bar D^*$, $D^* \bar D$, and $D^* \bar
D^*$, characterized by different values of the total spin $S_{tot}$.
The unpolarized total dissociation cross section comprises of
contribution $\sigma_f^{\rm unpol}(S_{tot})$ from different final states
$\{f\}$ and different total spins $\{S_{tot}\}$ of the system,
\begin{eqnarray}
\sigma_{tot}^{\rm unpol}=
\sum_f \, \sum_{S_{tot}} \sigma_f^{\rm unpol} (S_{tot}),
\end{eqnarray} 
where for this case with $L_A=1$ and $S_B=1$, $\sigma_f^{\rm unpol}
(S_{tot})$ is more complicated than the expression of Eq.\
(\ref{eq:unp1}) for $L_A=0$ and $S_B=1$.  It can be determined from
Eq.\ (\ref{eq:final}) and is given by
\begin{eqnarray}
\sigma_f^{\rm unpol} (S_{tot})
=
\sum_{ J\, M_A}
(\hat S \hat J)^2
\left \{
\matrix{  S_A   &    S_B   &  S  \cr
          L_A   &    0  &  L_A  \cr
          J_A   &    S_B   & J  \cr }
\right \}^2  
\sigma(L_A M_A S S_z),
\end{eqnarray}
where $\sigma(L_A M_A S S_z)$ is the cross section for the initial
meson system to have a total internal orbital angular momentum $L_A$
with azimuthal components $M_A$ and total spin $S$.  We show in Fig.\
13 the unpolarized total dissociation cross section for $\rho +\chi_0$
in Fig.\ 13$a$, $\rho +\chi_1$ in Fig.\ 13$b$, and $\rho +\chi_2$ in
Fig.\ 13$c$.  The exclusive cross section $\sigma_f^{\rm
unpol}(S_{tot})$ for different final states and $S_{tot}$ are also
shown.  The dissociation of $\chi_J$ in collision with $\rho$ is
exothermic.  The dissociation cross sections have the common features
that they diverge as $1/\sqrt{E_{KE}}$ near $E_{KE} \sim 0$ and
decreases monotonically as $E_{KE}$ increases.  The dominant
contribution to the dissociation cross sections comes from the $D^*
\bar D^* (S_{tot}=2)$ final state.  The unpolarized total dissociation
cross section for $\rho+\chi_0$ at $E_{KE}=0.05,$ 0.1, and 0.15 GeV
are 8.0 mb, 3.5 mb, and 1.6 mb respectively (Fig.\ 13$a$).  The
unpolarized total dissociation cross section for $\rho+\chi_1$ at
$E_{KE}=0.05,$ 0.1, and 0.15 GeV are 5.5 mb, 2.0 mb, and 0.8 mb
respectively (Fig.\ 13$b$).  The unpolarized total dissociation cross
section for $\rho+\chi_2$ at $E_{KE}=0.05,$ 0.1, and 0.15 GeV are 4.3
mb, 1.5 mb, and 0.5 mb respectively (Fig.\ 13$c$).  Thus, for the same
kinetic energy $E_{KE}$, $\sigma_{tot}^{\rm unpol}(\rho+\chi_0) >
\sigma_{tot}^{\rm unpol}(\rho+\chi_1) > \sigma_{tot}^{\rm
unpol}(\rho+\chi_2)$.

We show in Fig.\ 14 the unpolarized dissociation cross section of
$\chi_J$ in collision with $K$.  The lowest-energy final states are
$D_s\bar D^*$, $D_s^* \bar D$, and $D_s^* \bar D^*$.  For the
dissociation of $\chi_0$ in collision with $K$, the reaction has a
threshold at 0.07 GeV.  The total dissociation cross section rises
from the threshold to a maximum cross section of 1.7 mb at 0.27 GeV
(Fig.\ 14$a$).  For the dissociation of $\chi_1$ and $\chi_2$ in
collision with $K$, the reactions $K+\chi_1$ and $K+\chi_2$ are
exothermic for the final states of $D_s{\bar D}^*$ and $D_s^* \bar D$.
The total dissociation cross sections of $\chi_1$ and $\chi_2$ in
collision with $K$ behave as $1/\sqrt{E_{KE}}$ near $E_{KE} \sim 0$
but decrease very rapidly as $E_{KE}$ increases. These cross sections
then maintain a relatively constant value up to $E_{KE} \sim 0.2$ GeV
before decreasing at higher kinetic energies.

\vspace*{14cm}
\hskip -1cm
\epsfxsize=300pt
\includegraphics{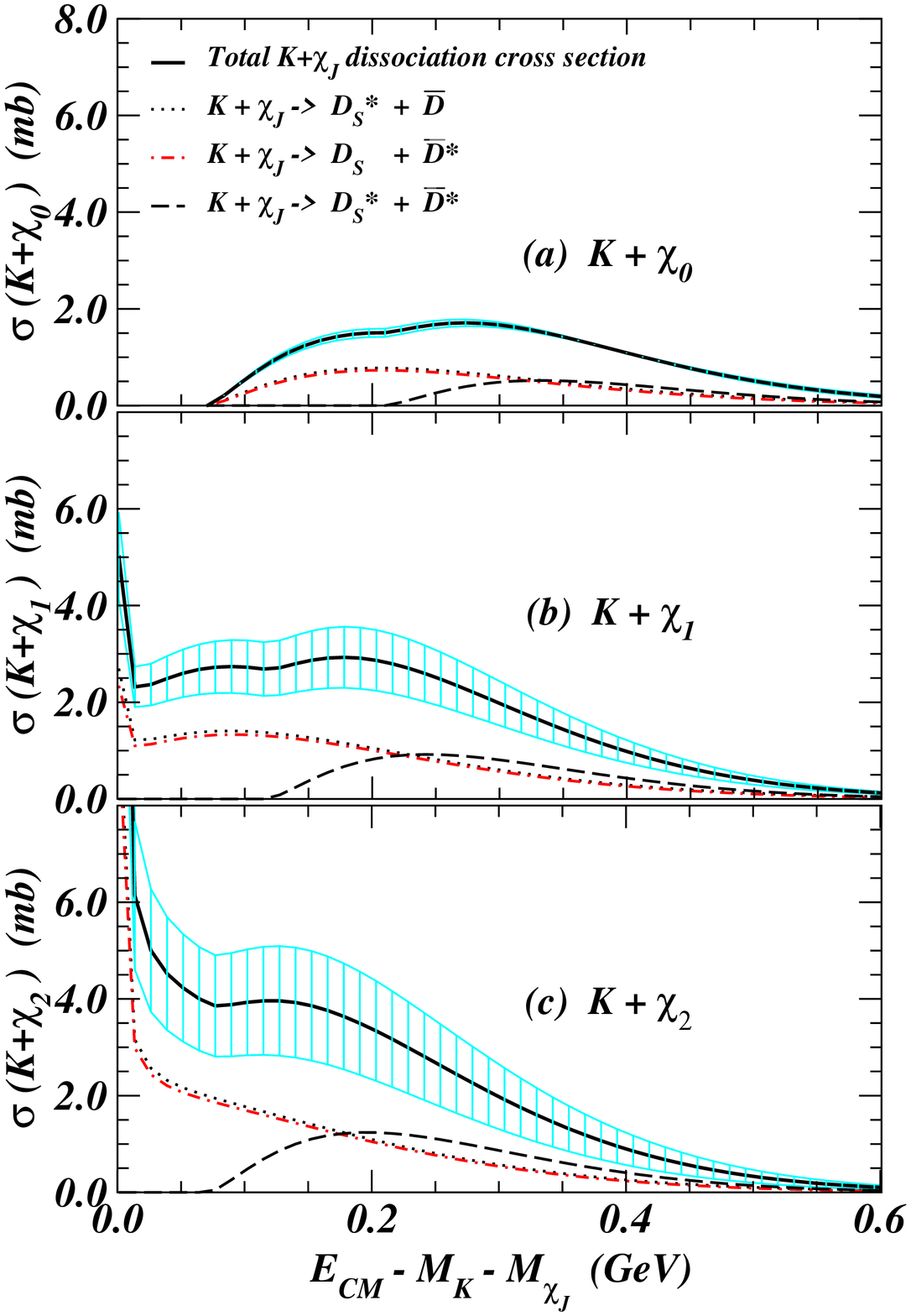}
\vspace*{0.5cm}\hspace*{1.5cm}
\begin{minipage}[t]{12cm}
\noindent {{\bf Fig.\ 14}.  Unpolarized cross sections for the
dissociation of $\chi_J$ in collision with $K$. Fig.\ 14$a$ is for
$K+\chi_0$, Fig.\ 12$b$ for $K+\chi_{1}$, and Fig.\ 14$c$ for
$K+\chi_{2}$.}
\end{minipage}
%\vskip 4truemm
\noindent 

\section{Discussion and Conclusions}

We have used the Barnes-Swanson quark-interchange model, with
parameters taken from fits to meson spectroscopy, to evaluate the
low-energy cross sections for the dissociation of the $J/\psi$,
$\psi'$, $\chi$, $\Upsilon$, and $\Upsilon'$ in collision with $\pi$,
$\rho$ and $K$.  The cross sections obtained here should be useful as
estimates of the importance of ``comover'' scattering in suppressing
heavy-quarkonium production, which is of considerable interest in the
search for the quark-gluon plasma.

Our results show that the cross section for the dissociation of
$J/\psi$ by $\pi$ occurs at a relatively high threshold, and the peak
total cross section is about 1 mb.  In contrast, the cross section for
the dissociation of $\psi'$ by $\pi$ occurs at a low threshold and the
cross section is much larger.  We have also evaluated the
corresponding cross section for the dissociation by $\rho$.  This
process is exothermic, and has a large total dissociation cross
section that diverges at threshold.

We previously noted that our $\pi$+$J/\psi$ cross section is
considerably smaller than the estimate of Ref.\cite{Mar95}, although
we use a similar approach.  There are several differences between the
two approaches which lead to this discrepancy.  First, Martins {\it et
al.}  assumed that the confining interaction is an attractive Gaussian
potential which acts only between quark-antiquark pairs.  The neglect
of the quark-quark and antiquark-antiquark confining interaction
amounts to discarding the transfer diagrams (T1 and T2) for the
confining potential. Since we find that the transfer and capture
diagram confinement contributions are similar in magnitude but
opposite in sign (due to color matrix elements), the confining
interaction assumed by Martins {\it et al.}  did not incorporate this
important destructive interference.  Second, their use of a Gaussian,
rather than the usual linear confining potential, obviously leads
to quantitatively different cross sections.  

The destructive interference between transfer and capture diagrams
with spin-independent forces (color-Coulomb and confinement) has been
noted previously. (See, for example, Refs.\cite{Bar92,Swa92} and
references cited in \cite{NN}.)  This interference explains the
well-known spin-spin hyperfine dominance in light hadron scattering in
channels such as $I=2$ $\pi\pi$, and the core $NN$ interaction. With
heavy quarks, however, the hyperfine interaction contribution is
smaller due to the large charm quark mass; for this reason we included
the color-Coulomb and confining interactions in our analysis.  Our
results indicates that the spin-spin, the linear interaction, and the
color-Coulomb interactions all give important contributions to the
dissociation cross sections.

It is of interest to compare our results of the dissociation cross
section with those obtained in the meson exchange model with effective
Lagrangians \cite{Mat98,Hag00,Lin00,Oh00}.  In the effective
Lagrangian approach, the dissociation cross section increases with
energy, as expected for the $t$-channel exchange of a spin-one
particle.  For example in Ref. \cite{Lin00} the dissociation cross
section is about 30 mb for $\pi + J/\psi\to D + \bar D^* $ and about
80 mb for $\pi + \Upsilon \to B + \bar B^*$, at 1 GeV above the
threshold.  These large cross sections continue to increase with
increasing energy.  In contrast, in our quark models calculation using
the Barnes-Swanson model, we find very small cross sections this far
above threshold for $\pi + J/\psi \to D + \bar D^* $ and \ $\pi +
\Upsilon \to B + \bar B^*$.  The predicated peak cross section for
$\pi + J/\psi \to D + \bar D^* $ is about 0.5 mb and occurs at about
0.05 GeV from the threshold.  The predicted peak cross section for
$\pi + \Upsilon \to B + \bar B^*$ is even smaller (about 0.03 mb), and
it occurs at about 0.02 GeV from the threshold.  These cross sections
decrease rapidly at higher energies.  Thus, the large cross sections
obtained in the effective Lagrangian approach differ by orders of
magnitude from the quark model results obtained here.  We believe that
the large increase in these dissociation cross sections predicted by
the effective Lagrangian meson exchange models at high energies is
unrealistic, since the momentum distributions of the boosted final and
the initial states have little overlap at high energies.

As the effective Lagrangian approach does not contain information
about the internal structure of the interacting hadrons,
phenomenological form factors have been introduced to reduce the
theoretical cross sections \cite{Hag00,Lin00,Oh00}.  A realistic
description of the form factors should incorporate the meson wave
functions and the dynamics of the scattering process.  Without a
derivation of these form factors, one encounters considerable
uncertainty, as experimental data on these reaction processes are
unavailable.  The form factors introduced in \cite{Hag00,Lin00,Oh00}
lead to changes of the theoretical cross section at high energies by
several orders of magnitude.  The results are sensitive both to the
assumed coupling strength and to the functional dependence of the form
factor.  In view of the strong dependence of the theoretical results
on the form factor and the coupling constants, a careful determination
of these quantities are required in future work.

Although there is no direct experimental measurement of these cross
sections to which we can compare our results, the small $\pi$+$J/\psi$
and the large $\pi$+$\psi'$ dissociation cross section at low kinetic
energies obtained here appear qualitatively consistent with earlier
results in a microscopic model of $J/\psi$ and $\psi'$ suppression in
O+A and S+U collisions \cite{Won96,Won98}.  Hopefully, future Monte
Carlo simulations of the dynamics of charmonium in heavy-ion
collisions will lead to a more direct comparison.  It is interesting
to note that dissociation of $J/\psi$ by $\pi$ and $\rho$ populate
different states (for example, $\pi$+$J/\psi$ does not lead to $D \bar
D$ in our model but $\rho$+$J/\psi$ does). It may provide a way to
separate these processes by studying the relative production of $D\bar
D$, $D^*\bar D$, $D {\bar D}^*$, and $D^*\bar D^*$, if the open charm
background can be subtracted.  However, the large ratio of initial
open charm to $J/\psi$ production in a nucleon-nucleon collision may
make this separation very difficult.

In the future it may be useful to carry out detailed simulations of
$J/\psi$ absorption in heavy-ion collisions using the cross sections
obtained here. If our cross sections do prove to be reasonably
accurate, it will clearly be useful to incorporate them in simulations
of hadron processes in relativistic heavy-ion collisions that use the
$J/\psi$ suppression as a signature of the quark-gluon plasma, in
order to isolate the effects of $J/\psi$ suppression due to its
interaction with hadron matter.

\section*{Acknowledgments}

This research was supported by the Division of Nuclear Physics,
Department of Energy, under Contract No. DE-AC05-00OR22725 managed by
UT-Battelle, LLC.  ES acknowledges support from the DOE under grant
DE-FG02-00ER41135 and DOE contract DE-AC05-84ER40150 under which the
Southeastern Universities Research Association operates the Thomas
Jefferson National Accelerator Facility.  The authors would also like
to thank Drs.\ C. M. Ko, H. Crater, and S. Sorensen for useful
discussions.

\appendix 
{\bf Appendix A:  Tabulation of Bound State Wave Functions}

The wavefunction in reduced momentum $2\bbox{p}=\bbox{p}_q - \bbox{
p}_{\bar q}$ is represented as a linear combination of Gaussian wave
functions with linearly-spaced $\beta^2$, of the form
$$
\Phi(2\bbox{p})
=\sum_{n=1}^N a_n \phi_n(2\bbox{p}) 
=\sum_{n=1}^N a_n
\left ( {1 \over \pi n \beta^2} \right ) ^ {3/4} 
{1 \over (2 n \beta^2)^{l/2}}
(2p)^l \sqrt{ {4 \pi \over {(2l+1)!!}} }Y_{lm}(\hat {\bbox{p}})
\exp \{ - { (2 \bbox{p})^2 \over 8 n \beta^2 } \}
$$
where $\Phi ({\bf2\bbox{p}})$ and $\phi_n (2{\bf\bbox{p}})$ are
normalized according to Eqs.\ (\ref{eq:Phinor}) and (\ref{eq:phinor}),
\begin{eqnarray}
\label{eq:Phi1}
\int d\bbox{p}~ |\Phi ({\bf2\bbox{p}})|^2 =
\int d\bbox{p}~ |\phi_n ({\bf2\bbox{p}})|^2 = 1.
\end{eqnarray}

The coefficients $\{ a_n\}$ for each meson in an $N$=6 basis, with a
different $\beta$ for each meson, are listed in the following table.

\hskip 2cm
\begin{tabular}{|c|c|c|c|c|c|c|c|c|c|} \hline
  Meson 	&$M$(exp)&$M$(th)&$\beta$ & $a_1$
&$a_2$
&$a_3$
&$a_4$
&$a_5$
&$a_6$ \\
                &(GeV)   &(GeV)  & (GeV)   &    &  &  &  &  &    \\
\hline
$\pi$            &   0.140 &   0.140 &   0.380
 &    0.8288 &   -0.5178 &   -0.2294 &    4.0001 &   -5.8837 &    2.9139 \\
\hline
$K$              &   0.494 &   0.495 &   0.440
 &    1.4258 &   -2.9104 &    6.6580 &   -7.6222 &    4.1972 &   -0.6622 \\
\hline
$K^*$            &   0.892 &   0.904 &   0.440
 &    2.6690 &   -7.7381 &   18.5854 &  -25.2611 &   17.6588 &   -4.9261 \\
\hline
$\rho$           &   0.770 &   0.774 &   0.380
 &    2.5214 &   -6.9921 &   16.7985 &  -22.9186 &   16.1163 &   -4.5409 \\
\hline
$\phi(1s)$       &   1.020 &   0.992 &   0.380
 &    1.4078 &   -2.2292 &    5.2488 &   -6.4976 &    4.0718 &   -0.9727 \\
\hline
$b_1$            &   1.235 &   1.330 &   0.380
 &    2.2568 &   -5.4759 &   12.6496 &  -17.1515 &   12.1528 &   -3.4443 \\
\hline
$a_1$            &   1.260 &   1.353 &   0.380
 &    2.3362 &   -5.7733 &   13.3524 &  -18.2221 &   13.0172 &   -3.7466 \\
\hline
$\phi(2s)$       &   1.686 &   1.870 &   0.380
 &    5.7964 &  -24.6635 &   58.1365 &  -79.9357 &   56.8725 &  -16.2428 \\
\hline
$D$              &   1.869 &   1.913 &   0.440
 &    1.8275 &   -4.2160 &   10.0225 &  -13.0384 &    8.6764 &   -2.2285 \\
\hline
$D^*$            &   2.010 &   1.998 &   0.440
 &    2.1630 &   -5.4765 &   13.0711 &  -17.5068 &   12.0520 &   -3.2893 \\
\hline
$D_s$            &   1.969 &   2.000 &   0.440
 &    1.0701 &   -1.1418 &    2.4522 &   -1.9688 &    0.3196 &    0.3292 \\
\hline
$D_s^*$          &   2.112 &   2.072 &   0.440
 &    1.3267 &   -1.9616 &    4.5086 &   -5.2172 &    2.9478 &   -0.5646 \\
\hline
$D_1(^1P_1)$     &   2.422 &   2.506 &   0.440
 &    2.2042 &   -5.2226 &   12.0872 &  -16.4595 &   11.7325 &   -3.3595 \\
\hline
$D_2(^3P_2)$     &   2.460 &   2.514 &   0.440
 &    2.2344 &   -5.3296 &   12.3210 &  -16.7896 &   11.9756 &   -3.4375 \\
\hline
$\eta_c$         &   2.979 &   3.033 &   0.560
 &    0.9461 &   -0.6474 &    1.0666 &    0.3614 &   -1.6509 &    0.9868 \\
\hline
$J/\psi$         &   3.097 &   3.069 &   0.560
 &    1.0786 &   -1.0517 &    2.0729 &   -1.2289 &   -0.3804 &    0.5646 \\
\hline
$h_c$            &   3.570 &   3.462 &   0.560
 &    1.6312 &   -2.8587 &    6.7068 &   -9.0601 &    6.4161 &   -1.8163 \\
\hline
$\chi_c$         &   3.525 &   3.466 &   0.560
 &    1.6587 &   -2.9420 &    6.8805 &   -9.3051 &    6.5918 &   -1.8698 \\
\hline
$\psi'$          &   3.686 &   3.693 &   0.560
 &    5.5237 &  -22.5889 &   53.5145 &  -74.6754 &   53.9666 &  -15.7222 \\
\hline
$B$              &   5.279 &   5.322 &   0.500
 &    2.4905 &   -7.0584 &   17.0138 &  -23.0809 &   16.1128 &   -4.4690 \\
\hline
$B^*$            &   5.324 &   5.342 &   0.500
 &    2.5806 &   -7.4190 &   17.8694 &  -24.2889 &   16.9859 &   -4.7275 \\
\hline
$B_s$            &   5.369 &   5.379 &   0.500
 &    1.6289 &   -3.2614 &    7.6922 &   -9.8055 &    6.3702 &   -1.5866 \\
\hline
$B_s^*$          &   5.416 &   5.396 &   0.500
 &    1.7111 &   -3.5483 &    8.3718 &  -10.7991 &    7.1136 &   -1.8177 \\
\hline
$\Upsilon(1s)$   &   9.460 &   9.495 &   0.660
 &    0.1364 &    2.0441 &   -6.7818 &   14.2875 &  -13.9803 &    5.3693 \\
\hline
$\chi_b(1p)$     &   9.899 &   9.830 &   0.660
 &    0.7416 &    0.1481 &    0.2178 &   -0.0196 &   -0.1923 &    0.1587 \\
\hline
$\Upsilon(2s)$   &  10.020 &   9.944 &   0.660
 &   -3.6422 &   10.7655 &  -25.3407 &   38.7395 &  -30.5647 &    9.8544 \\
\hline
$\chi_b(2p)$     &  10.260 &  10.166 &   0.660
 &    3.2645 &  -10.1170 &   20.7960 &  -28.2373 &   20.2666 &   -5.9472 \\
\hline
\end{tabular}


\begin{thebibliography}{99}
\bibitem{QM99} {\it Quark Matter '99}, Proceedings of the 14th
International Conference on Ultra-Relativistic Nucleus-Nucleus
Collisions, Editors L. Riccati, M. Masera, and E. Vercellin, Nuclear
Physics A, Volume A661 (1999).

\bibitem{Won94}
C. Y. Wong, {\it Introduction to High-Energy Heavy-Ion Collisions},
World Scientific Publishing Company, 1994.

\bibitem{Hei00}
U. Heinz and M. Jacobs, nucl-th/0002042.

\bibitem{Won01}
C. Y. Wong, Nucl. Phys. {\bf A681}, 22c (2001).

\bibitem{Mat86} T. Matsui and H. Satz, Phys. Lett. {\bf B178}, 416
(1986).

\bibitem{Gon96} M. Gonin, NA50 Collaboration, Nucl. Phys. 
{\bf A610}, 404c (1996). 

\bibitem{Rom98} A. Romana $et~al.$, NA50 Collaboration, in Proceedings
of the XXXIII Recontres de Moriond, Les Arcs, France, 21-28 March,
1998.

\bibitem{Won96} C. Y. Wong, Nucl. Phys.  {\bf A610}, 434c (1996);
Nucl.Phys. {\bf A630}, 487 (1998).

\bibitem{Won98} C. Y. Wong, hep-ph/9809497, in proceedings of Workshop
on Charmonium Production in Relativistic Nuclear Collisions, (INT,
Seattle, May 18-22, 1998).

\bibitem{Kha96} D. Kharzeev, Nucl. Phys.  {\bf A610}, 418c (1996);
Nucl. Phys. {\bf A638}, 279c (1998).

\bibitem{Bla96} J.-P. Blazoit and J.-Y. Ollitrault, 
Nucl. Phys.  {\bf A610}, 452c  (1996).

\bibitem{Cap96} A. Capella, A. Kaidalov, A. K. Akil, and C. Gerschel,
Phys. Lett. {\bf B393}, 431  (1997).

\bibitem{Cas96} W. Cassing and C. M. Ko, Phys. Lett. {\bf B396}, 39
 (1997); W. Cassing, E. L. Bratkovskaya, Nucl. Phys. {\bf A623}, 570
 (1997).

\bibitem{Vog98}
R. Vogt, Phys. Lett. {\bf B430} 15 (1998).

\bibitem{Nar98} M. Nardi and H. Satz, Phys. Lett. {\bf B442} 14
(1998).

\bibitem{Zha00} Bin Zhang, C.M. Ko, Bao-An Li, Ziwei
Lin, and Ben-Hao Sa, Phys.Rev. {\bf C62}, 054905  (2000).

\bibitem{Bla00}
D.B. Blaschke, G. R. G. Burau, M. I. Ivanov, Yu. L. Kalinovsky, and
P. C. Tandy, Phys. Lett. {\bf B506} 297 92001).

\bibitem{Kha94} D. Kharzeev and H. Satz, Phys. Lett. {\bf B334}, 155
(1994).

\bibitem{Kha96a} D. Kharzeev, H. Satz, A. Syamtomov, and G. Zinovjev,
Phys. Lett. {\bf B389}, 595 (1996).

\bibitem{Mar95}
K. Martins, D. Blaschke, and E.\ Quack, Phys.\ Rev. {\bf C51}, 2723 (1995).

\bibitem{Won00a}
C. Y. Wong, E. S. Swanson, and T. Barnes,
Phys. Rev. {\bf C62}, 045201, 2000. 

\bibitem{Won00b}
C. Y. Wong, E. S. Swanson, T. Barnes, in Proceedings of the 
Hirschegg 2000 Workshop on Hadrons in Dense Matter
(Hirschegg, Austria, 16-22 January 2000);
nucl-th/000203.

\bibitem{Bar00} T.Barnes, E.S.Swanson, and C.Y.Wong, Proceedings of
International Workshop on Heavy Quark Physics 5 (Dubna, 6-8 April
2000), nucl-th/0006012.

\bibitem{Mat98}
S. G. Matinyan and B. M\"uller, Phys. Rev. {\bf C58}, 2994 (1998).

\bibitem{Hag00}
K. L. Haglin, Phys. Rev. {\bf C61}, 031912 (2000);
K. L. Haglin and C. Gale, Phys. Rev. {\bf C63} 065201 (2001).

\bibitem{Lin00}
Z. W. Lin and C. M. Ko, 
Phys. Rev. {\bf C62}, 034903 (2000).

\bibitem{Oh00} Yongseok Oh, Taesoo Song, and Su Houng Lee,
Phys. Rev. {\bf C63}, 034901 (2001).

\bibitem{Sib00}
A. Sibirtsev, K. Tsushima, and A. W. Thomas, 
Phys. Rev. {\bf C63}, 044906  (2001).

\bibitem{Pes79} M. Peskin, Nucl. Phy. {\bf B156}, 365 (1979).

\bibitem{Bha79}
G. Bhanot and M. Peskin, Nucl. Phy. {\bf B156}, 391 (1979).

\bibitem{Ant93}
L. Antoniazzi $et~al.$, Phys. Rev. Lett. {\bf 70}, 383 (1993).

\bibitem{Lin01} Ziwei Lin and C. M. Ko, Phys. Lett. {\bf B503}, 104
2001.

\bibitem{Bar92}
T. Barnes and E. S. Swanson, Phys. Rev. {\bf D46}, 131 (1992).

\bibitem{Swa92}
E. S. Swanson, Ann. Phys. (N.Y.) {\bf 220}, 73 (1992).

\bibitem{Kpi}
T. Barnes, E. S. Swanson, and J. Weinstein, Phys. Rev. {\bf D46}, 4868 (1992).

\bibitem{KN}
T. Barnes and E. S. Swanson, Phys. Rev. {\bf C49}, 1166 (1994).

\bibitem{NN} T. Barnes, S. Capstick, M. D. Kovarik, and E. S. Swanson,
Phys. Rev. {\bf C48}, 539 (1993).

\bibitem{Bar99}
T.\ Barnes, N.\ Black, D.\ J.\ Dean, and E.\ S.\ Swanson,
Phy. Rev. {\bf C60}, 045202 (1999).

\bibitem{Sch68}
L. I. Schiff, {\it Quantum Mechanics} (McGraw-Hill, New York, 1968), 
pp.384-387. 

\bibitem{Tal63}
A. de-Shalit and I. Talmi,
{\it Nuclear Shell Model}, Academic Press, 1963.

\bibitem{Hoo77}
W. Hoogland $et~al.$, Nucl. Phys. B{\bf 126}, 109 (1977).

\bibitem{Won00c}
C.Y. Wong and H. Crater,  Phys. Rev. {\bf C63}, 044907 (2001).

\bibitem{Bar01}
T. Barnes, E. Swanson, and C.Y. Wong, to be published.

\end{thebibliography}
\end{document}